\documentclass[12pt]{article}

\input epsf
\usepackage{amssymb,amsmath,wrapfig,subfigure}
\usepackage[matrix,arrow,curve]{xy}
\usepackage{cite}
\usepackage{graphicx,color}
\usepackage[debug,pageanchor=false]{hyperref}
\definecolor{link}{rgb}{.8,.15,.1}
\hypersetup{colorlinks=true,linkcolor=link,citecolor=link,urlcolor=link,linktocpage}

\makeatletter
\@addtoreset{equation}{section}
\makeatother

\def\rr {{\Bbb R}}
\def\cc {{\Bbb C}}
\def\pp {{\Bbb P}}
\def\zz {{\Bbb Z}}
\def\del {\partial}

\def\del {\partial}

\def\vol {\mathrm{vol}}

\begin{document}

\begin{titlepage}

\begin{center}

\vskip .3in \noindent

{\Large \bf{Supersymmetry on Three-dimensional Lorentzian \\\vspace{.2cm} Curved Spaces and Black Hole Holography}}

\bigskip

	Kiril Hristov, Alessandro Tomasiello and Alberto Zaffaroni\\

       \bigskip
		 Dipartimento di Fisica, Universit\`a di Milano--Bicocca, I-20126 Milano, Italy\\
       and\\
       INFN, sezione di Milano--Bicocca,
       I-20126 Milano, Italy

       \vskip .5in
       {\bf Abstract }
       \vskip .1in

       \end{center}

We study ${\cal N}\le 2$ superconformal and supersymmetric theories on Lorentzian three-manifolds with a view toward holographic applications, in particular to BPS black hole solutions. As in the Euclidean case, preserved supersymmetry for asymptotically locally AdS solutions implies the existence of a (charged) ``conformal Killing spinor'' on the boundary. We find that such spinors exist whenever there is a conformal Killing vector which is null or timelike. We match these results with expectations from supersymmetric four-dimensional asymptotically AdS black holes. In particular, BPS bulk solutions in global AdS are known to fall in two classes, depending on their graviphoton magnetic charge, and we reproduce this dichotomy from the boundary perspective. We finish by sketching a proposal to find the dual superconformal quantum mechanics on the horizon of the magnetic black holes.
\noindent

\vfill
\eject

\end{titlepage}

\section{Introduction} 
\label{sec:intro}

The subject of supersymmetric field theories on curved spaces has found a wide range of applications in recent years, see for example  \cite{Pestun:2007rz,Kapustin:2009kz,Jafferis:2010un,Hama:2010av,Hama:2011ea,Imamura:2011wg}. 
The construction of such theories was shown to follow from an algorithmic procedure \cite{Festuccia:2011ws} with off-shell supergravity as a starting point, treating the gravity multiplet fields as non-dynamic curved background for the field theory. This approach has led to a classification of supersymmetric backgrounds for ${\cal N} = 1$ theories in four dimensions with Euclidean and Lorentzian signature \cite{ Klare:2012gn,Dumitrescu:2012ha,Cassani:2012ri,Dumitrescu:2012at}, for ${\cal N}=2$ theories in three Euclidean dimensions \cite{Klare:2012gn,Closset:2012ru} and to many other results in various signatures and dimensions  \cite{Jia:2011hw,Samtleben:2012gy,Liu:2012bi,deMedeiros:2012sb,Kehagias:2012fh, Samtleben:2012ua,Kuzenko:2012vd}.
The cases when the supersymmetric field theories are additionally conformally invariant \cite{Klare:2012gn} were then found to follow from freezing the gravity multiplet in conformal supergravity 
\cite{Kaku:1977rk,Kaku:1977pa,Kaku:1978nz,Ferrara:1977mv}.  These superconformal field theories (SCFTs) deserve special attention due to the possibility of a holographic description of such theories in terms of a bulk gravity dual. From both the conformal supergravity and holographic approaches, it was found in \cite{Klare:2012gn,Cassani:2012ri} (for the Euclidean three- and four-dimensional case, and Lorentzian four-dimensional case, respectively) that a SCFT retains some supercharges on a curved space only if it admits a solution to the conformal Killing spinor (CKS) equation. 

In this paper, we turn our attention to three-dimensional ${\cal N}=2 $  Lorentzian superconformal theories and their relevance for describing four-dimensional bulk duals. In this case as well the spacetime needs to allow for a conformal CKS solution, possibly charged under a gauge field, in order to admit a consistent SCFT. 
We will characterize the existence of a CKS in terms of the existence of a conformal Killing vector (CKV). This is similar to what was found for the Lorentzian four-dimensional case in \cite{Cassani:2012ri}, except that in that case the CKV needed to be null, whereas in our case we will see that it can be timelike or null. The timelike case corresponds to a CKS which is generically charged under a gauge field, which can be computed from geometric data. The null case corresponds to a CKS which is Majorana and uncharged. 

Apart from their purely field theoretic interest, three-dimensional SCFTs are related via holography to four-dimensional quantum gravity; in this paper, we explore in particular the link with black holes. In this introduction, let us for example focus on static solutions whose asymptotic geometry is global AdS$_4$.\footnote{Here we mean solutions whose conformal boundary has $\rr\times S^2$ topology. It is also possible to quotient the conformal boundary in a certain way, and obtain horizons with topology $T^2$ or higher-genus Riemann surfaces (see \cite{Caldarelli:1998hg} for details). Although most of our interest will be in spherical horizons, later we will consider these more exotic cases as well.} There are two known classes of such solutions
; both of these were originally found in \cite{Romans:1991nq} in minimal supergravity. Solutions in the first class are $1/2$ BPS; all known ones (even in non-minimal supergravity) have naked singularities. Solutions in the second class are $1/4$ BPS; recently, some examples with finite-area horizon have been found in \cite{Cacciatori:2009iz} (and elaborated upon in \cite{Dall'Agata:2010gj,Hristov:2010ri}). The graviphoton field strength has a different asymptotic behavior for these two types of solutions: 
\begin{equation}\label{eq:dic}
	F \to 0  \qquad (1/2 \ {\rm BPS})\,; \qquad F \to -\frac12 {\rm vol}_{S^2} \qquad (1/4\ {\rm BPS})\,,
\end{equation}
where $S^2$ is a spatial slice of global AdS. In the second case, notice that the integral of the field-strength is $\int_{S^2} F = -2\pi$, the smallest negative value allowed by flux quantization. For this reason, the asymptotic structure of these solutions was called ``magnetic AdS$_4$'' in \cite{Hristov:2011ye}. 

In this paper, we reproduce the dichotomy (\ref{eq:dic}) by classifying explicitly all CKVs of the conformal boundary $\rr\times S^2$. We find that there are only two rotationally invariant gauge fields which admit a CKS: $F=0$ (corresponding to the first class, the $1/2$ BPS solutions) and $F=-\frac12{\rm vol}_{S^2}$ (corresponding to the second class, the $1/4$ BPS black holes).

This encouraging result opens the door to more ambitious questions. Perhaps the most interesting issue about quantum black holes is their microscopic entropy counting. 
Starting from the first successful instance of such counting for certain asymptotically flat BPS black holes in \cite{Strominger:1996sh,Maldacena:1997de}, the understanding of black holes has led to many revelations and ongoing debates about the nature of quantum gravity. It is important to find a microscopic description for more general classes of black holes. One approach consists in applying holography to the near-horizon geometry: this contains an AdS$_2$ factor, and from the AdS$_2$/CFT$_1$ correspondence one expects to find a one-dimensional superconformal quantum mechanics capturing the microscopic degrees of freedom (see for example \cite{Claus:1998ts,BrittoPacumio:1999ax,Gaiotto:2004ij,Gaiotto:2004pc,Sen:2008vm}). Usually there is no systematic way to obtain the correct CFT, whose states reproduce the black hole entropy. For black holes in AdS, however, one can also use holography at the boundary. As we mentioned earlier, \cite{Cacciatori:2009iz} found static 1/4 BPS black holes which have AdS$_4$ asymptotics at infinity, and ${\rm AdS}_2 \times S^2$ in the near-horizon limit. Moreover, these black hole solutions can be embedded in string theory, for example in ${\rm AdS}_4\times S^7$ in M-theory. This gives a strategy to obtain the CFT$_1$. One can consider the CFT$_3$ dual to the AdS$_4$ asymptotical background (for example ABJM \cite{abjm}), put it on $\rr\times S^2$ and deform it in a way dual to the behavior at infinity of the scalar fields. The renormalization group (RG) flow of such a theory should be holographically dual to the black hole solution. In the infrared, the CFT$_3$ should flow to a CFT$_1$ dual to the ${\rm AdS}_2\times S^2$ near-horizon geometry. In this paper we will perform some checks on this overall picture, by computing the relevant superconformal algebras.
In section \ref{sec:cks} we argue that charged CKS are necessary in order for an ${\cal N}=2$  SCFT to preserve some supersymmetry, and prove that they exist if and only if a null or timelike CKV exists. In section \ref{sec:coord} we introduce local coordinates, distinguishing between two main classes of metrics: the ones admitting a lightlike CKV and the ones admitting a timelike CKV, thus matching the expectation from the four-dimensional bulk classification of supersymmetric backgrounds in \cite{Caldarelli:2003pb,Cacciatori:2004rt}. In section \ref{sec:examples} we spell out more explicitly what our general formalism tells us about CKS solutions for conformally flat metrics, showing that one can generically define several unrelated SCFTs on such manifolds. We discuss in some detail $\rr\times S^2$, Minkowski$_3$, and $\rr\times H_2$, which are of particular relevance for black hole holography. In section \ref{sec:bh} we give a brief overview of the bulk black hole solutions in four dimensions that we try to describe holographically. We then give a prescription to mass deform and dimensionally reduce a three-dimensional SCFT to a one-dimensional superconformal quantum mechanics, thus relating the dual theory on the boundary at infinity to the dual theory on the horizon of a four-dimensional black hole. We elaborate on this connection in detail for the particular cases of $\rr \times S^2$ with a background magnetic field and $\rr \times T^2$ with a vanishing magnetic field. These two cases correspond exactly to the asymptotically magnetic AdS and Riemann AdS bulk geometries \cite{Hristov:2011ye,Hristov:2011qr,Hristov:2012bd}. We then show more briefly the analogous results for higher genus black holes in AdS. 

In appendix \ref{sec:conf} we also make some general comments (valid in any dimension, not just three) about the CKV and CKS equations and their conformal invariance; we illustrate our comments by considering SCFTs on de Sitter space.


\section{Geometry of conformal Killing spinors in three dimensions} 
\label{sec:cks}

In this section, we will consider the properties of (charged) conformal Killing spinors in three dimensions, namely solutions to
\begin{equation}\label{eq:cks}
	\left(\nabla^A_\mu - \frac13 \gamma_\mu D^A \right) \epsilon = 0 \ ,
\end{equation}
where $\nabla^A\equiv \nabla - i A$ and $D^A \equiv \gamma^\mu \nabla^A_\mu$ is the Dirac operator. In section \ref{sub:qft} we will  argue that they are necessary in order for a three-dimensional SCFT to preserve some supersymmetry  on a curved space. In the rest of the section we   will discuss the geometry of the CKS.

\subsection{Three-dimensional SCFTs on curved space} 
\label{sub:qft}

A general strategy for defining a SCFT theory on curved space  is to couple it to conformal supergravity and to freeze the gravitational 
fields to some background value, in analogy with the suggestion of  \cite{Festuccia:2011ws} for supersymmetric theories. Supersymmetry will be preserved if the  variation of the  gravitino vanishes. The strategy has been used in  \cite{ Klare:2012gn,Dumitrescu:2012ha,Cassani:2012ri,Dumitrescu:2012at} for studying four-dimensional  and Euclidean three-dimensional SCFTs and we refer to
those papers for more details. In this article we  will focus on theories with ${\cal N}= 2$ supersymmetry in three dimensions. 

The  ${\cal N} = 2$  conformal supergravity multiplet in three dimensions  consists of a graviton $g_{\mu\nu}$,   a Dirac
gravitino $\psi_\mu$ and a real gauge field $A_\mu$. The gravitino variation is given by \cite[Sec.~3]{Rocek:1985bk}
\begin{equation}\label{eq:gravitinoN2}
\delta \psi_\mu   =	\nabla^A_\mu \epsilon + \gamma_\mu \eta  \ ,
\end{equation}
where $\epsilon$ is the parameter for the supersymmetries and $\eta$ is the parameter for the superconformal transformations; $\epsilon$ and $\eta$ are Dirac spinors.  We will be able to define an  ${\cal N} = 2$ SCFT on a curved manifold $M_3$ if and only if
the right hand side of the previous equation vanishes from some $\epsilon$ and $\eta$. By taking the gamma trace of  the resulting condition 
\begin{equation}
\label{eq:superconf}
\nabla^A_\mu \epsilon =-\gamma_\mu \eta
\end{equation}
we can  eliminate $\eta$  and we recover equation (\ref{eq:cks}).  We thus conclude that in order for three-dimensional SCFT to preserve some supersymmetry  on a curved space we need the existence of a charged CKS \footnote{We notice that our results will have applications also to the problem of defining  supersymmetric (but not necessarily conformal) theories with an R-symmetry  on curved spaces. Indeed, as noticed in \cite{Klare:2012gn}, the solutions of the CKS equation are closely related to the  solutions of the supersymmetry conditions of new minimal supergravity. We refer to  \cite[Sec.~5]{Klare:2012gn} for a discussion of this relation in three dimensions and to \cite{Klare:2012gn,Closset:2012ru} for examples in Euclidean signature.}. 

The Lagrangian of a SCFT in curved space can be explicitly written by considering the coupling to  conformal supergravity and by freezing the metric and the gauge field $A$ to the values that admit a solution of (\ref{eq:cks}). Matter couplings in extended supergravity theories in three dimensions are less known than in four dimensions but they have been recently discussed in superspace in \cite{Kuzenko:2011xg,Kuzenko:2011rd,Kuzenko:2012ew}. At linearized level the Lagrangian for flat space is corrected by the terms
\begin{equation}\label{eq:cscoupling}
	-\frac 12 g_{\mu\nu}T^{\mu\nu} + A_\mu J^\mu + \bar \psi_\mu {\cal J}^\mu \ ,
\end{equation}
where $J^\mu$ is the R-symmetry and ${\cal J}^\mu$ is the supersymmetry current of the SCFT. In particular $A_\mu$ appears
as a background gauge field for the $U(1)_R$ symmetry of the SCFT.

It is interesting to notice that equation (\ref{eq:cks}) can be also derived from a holographic perspective \cite{Klare:2012gn}.  We start 
with an ${\cal N}=2$ gauged supergravity with an ${\rm AdS}_4$ vacuum corresponding to the dual  of a three-dimensional superconformal field theory on flat space. We can consider the three-dimensional  fields $g_{\mu\nu},\psi_\mu,A_\mu$ as boundary values of the
bulk fields of the gauged supergravity. Indeed, according to the principles of the AdS/CFT correspondence, the bulk fields couple to
CFT operators through their boundary coupling as in equation (\ref{eq:cscoupling}). The dual of the CFT on a curved
manifold $M_3$ with a non trivial background field $A_\mu(x)$ is described by a solution of four-dimensional gauged supergravity whose metric is asymptotically locally AdS (AlAdS), namely such that
\begin{equation}\label{asymp}
		ds_4^2 =  \frac{dr^2}{r^2} + \left ( r^2 ds_{M_3}^2 + O(r) \right ) \ , \\
\end{equation}
and whose graviphoton behaves as
\begin{equation}\label{asympA}
	A_\mu  = A_\mu (x) + O\left(\frac1r\right) \ , \qquad  \mu=0,1,2\ .
\end{equation}
We can derive the boundary supersymmetry conditions on $M_3$ and $A$  by
expanding the bulk supersymmetry conditions for large $r$.  The analysis is identical to the  one discussed in  \cite[Sec.~2.1]
{Klare:2012gn} for the Euclidean case, and it will not be repeated here. The result is that the bulk supersymmetry parameter
behaves near the boundary as
\begin{equation}
\epsilon_{4d} = r^{1/2} \tilde \epsilon + r^{-1/2} \tilde \eta
\end{equation}
where $\tilde \epsilon$ and $\tilde\eta$  reduce to a pair of three-dimensional spinors $\epsilon$ and $\eta$ satisfying equation (\ref{eq:superconf}), which is in turn equivalent to the existence of a CKS.

Finally, the case of ${\cal N}=1$ supersymmetry can be considered as a special case of the previous discussion. The ${\cal N}=1$ conformal 
supergravity contains a graviton  and a Majorana gravitino $\psi_\mu$. The supersymmetry spinor $\epsilon$ is Majorana. 
We thus conclude that in order for an ${\cal N}=1$  three-dimensional SCFT to preserve some supersymmetry  on a curved space we need the existence of a (uncharged) Majorana CKS. 

\subsection{Algebraic constraints} 
\label{sub:alg}

In this section we will  review the geometry behind a spinor $\epsilon$ in Lorentzian three dimensions. We will generically take $\epsilon$ to be Dirac, and see the Majorana condition as a special case. To fix ideas, we will work in a basis where all gamma matrices are real (e.g. $\gamma^0=i \sigma_2$, $\gamma^1=\sigma_1$, $\gamma^2= \sigma_3$).

As usual, we can start with the bispinor $\epsilon \otimes \overline{\epsilon}$, where $\overline{\epsilon}\equiv \epsilon^\dagger \gamma^0$.  We can reexpress it in terms of forms using the Fierz identities. In odd dimensions, the slash (or Clifford map) of a $k$-form is proportional to the slash of a $(d-k)$-form, and so a basis for bispinors can be obtained by considering not all the forms but half of them. In three dimensions, for example, $1$ and $\gamma_\mu$ (the slashes of zero- and one-forms) can be used as a basis: 
\begin{equation}\label{eq:qz}
	\epsilon \otimes \overline{\epsilon} = i \beta + z\ ,
\end{equation}
where $z= z_\mu \gamma^\mu$, and
\begin{equation}\label{eq:betaz}
	i \beta = \frac12 \overline{\epsilon} \epsilon \ ,\qquad 
	z_\mu = \frac12 \overline{\epsilon} \gamma_\mu \epsilon\ .
\end{equation}
$\beta$ and $z$ are real. 

Computing the action of $z$ on $\epsilon$, we find
\begin{equation}\label{eq:ze}
	z \epsilon= \frac12 \gamma^\mu \epsilon \overline{\epsilon} \gamma_\mu \epsilon = \frac12 (3i \beta-z) \epsilon \quad \Rightarrow \quad z \epsilon = i \beta \epsilon \quad \Rightarrow \quad z^2 = - \beta^2\ .
\end{equation}
where we have used $\gamma^\mu C_k \gamma_\mu = (-1)^k (3-2 k) C_k$, for $C_k$ the slash of a $k$-form.


\subsection{Spinors and vectors} 
\label{sub:sv}

So the vector $z$ can be either null or timelike. In our basis, where $\gamma^0= i \sigma_2$, it is immediate to see that
\begin{equation}
	\beta=0 \quad \Rightarrow \qquad \epsilon= e^{i \alpha} \epsilon_0 \ \qquad (\epsilon_0\ {\rm Majorana}). 
\end{equation}

We have seen that $\epsilon$ determines a vector $z$; for our purposes, it will be important to understand to what extent we can invert this map. To simplify the discussion, let us consider spinors $\epsilon$ normalized such that $\epsilon^\dagger \epsilon=1$. The space of such spinors is an $S^3$. Since $\epsilon^\dagger \epsilon= \overline{\epsilon}\gamma_0 \epsilon = z_0$, this sphere is mapped to the space of vectors $z$ which are null or timelike and with $z_0=1$: namely, a disk $D$. So we have a map 
\begin{equation}
 \{ \epsilon \ | \ \epsilon^\dagger \epsilon=1\}= S^3\ \longrightarrow \ D = \{ z \ | \ z_0=1, z^2\le 0 \}\ . 
\end{equation}
The counterimage of a $z$ on the boundary of $D$ is a $S^1$: it consists of spinors of the form $e^{i \alpha} \epsilon_0$, where $\epsilon_0$ Majorana is determined uniquely. On the other hand, the counterimage of a $z$ in the interior of $D$ consists of two $S^1$s. Indeed, in this case, given an $\epsilon$ whose image is a given timilike $z$, not only all spinors of the form $e^{i \alpha}\epsilon$ are mapped to $z$, but also all spinors of the form $e^{i \alpha} \epsilon^*$. These two $S^1$s can also be seen in a different way: let $\epsilon= \epsilon_1 \pm i\epsilon_2$, with $\epsilon_{1,2}$ Majorana. If we call $z_{1,2}$ the vector associated to $\epsilon_{1,2}$, then $z=z_1+z_2$, for a given $z$ we need to find two null vectors $z_1$, $z_2$ such that $z= z_1 + z_2$. This can be done in a $S^1$ worth of ways (think for example about the case $z=(1,0,0)$). Again, since $\epsilon= \epsilon_1 \pm i \epsilon_2$, there are two $S^1$s. 

As a cross-check, if we reconstruct the fibration from these counterimages, we see that we have two copies of a solid torus $D\times S^1$ (corresponding to the two $S^1$s we just discussed) glued about the boundary in a certain way; it is well known that $S^3$ can arise in this way. 

We have seen that a spinor $\epsilon$ can be reconstructed pointwise from its vector bilinear $z_\mu = \frac12 \overline{\epsilon} \gamma_\mu \epsilon$, up to a U(1) and, when $z$ is timelike, up to a discrete $\zz_2$ choice (whose origin is conjugation). A priori this discrete ambiguity exists for every point, but, if we assume $z$ to be continuous, fixing it at one point will fix it everywhere.\footnote{If there are codimension-one loci where $z$ is timelike, the ambiguity can be fixed in two different ways on each side of the locus.}
Up to this global discrete ambiguity, the vector field $z$ determines the spinor $\epsilon$ up to the U(1) gauge invariance 
\begin{equation}
	\epsilon \to e^{i \alpha} \epsilon\ .
\end{equation}
This will be the ${\rm U}(1)$ gauge invariance of the gauge field $A$ in (\ref{eq:cks}).



\subsection{$G$-structures} 
\label{sub:gstr}

The null and timelike case define two different $G$-structures. These can be computed simply as the stabilizer of the null vector $z$, since it is the only form bilinear that can be defined from $\epsilon$. ($\beta$ in (\ref{eq:ze}) is a zero-form and does not impose any conditions on a stabilizer.) The stabilizer of a null vector in $d$ dimensions is ${\rm SO}(d-2)\ltimes \rr^{d-2}$, so in $d=3$ a null vector has a stabilizer $\rr$. On the other hand, the stabilizer of a timelike vector is clearly ${\rm SO}(2)$.

One could treat the two different cases together as follows. Along the lines of \cite[Sec.~3.1]{Cassani:2012ri}, one can introduce a one-form $e^-$ whose function is to break both $\rr$ and ${\rm SO}(2)$ to the identity, so that one now has an identity structure, or in other words a vielbein. We can demand $e^-$ to satisfy
\begin{equation}\label{eq:e-}
	(e^-)^2=0 \ ,\qquad z\cdot e^- = 2 \ .
\end{equation}
If one introduces now 
\begin{equation}
	e^2_\mu \equiv \frac12 \epsilon_{\mu\nu\rho} z^\nu e^{-\rho} \ ,\qquad e^+_\mu = z_\mu + \frac14 \beta^2 e^-_\mu \ , 
\end{equation}
one has 
\begin{equation}
	e^+\cdot e^-= 2 \ ,\qquad e^2 \cdot e^\pm=0 \ ,
\end{equation}
which indeed together with (\ref{eq:e-}) tell us that $\{ e^+, e^-, e^2\}$ are a vielbein. One also has a basis of spinors:
\begin{equation}\label{eq:e-basis}
	\epsilon \ ,\qquad e^- \epsilon \ .
\end{equation}

The drawback of this strategy is that it is computationally heavy: for example, the action of $\gamma_\mu$ in the basis (\ref{eq:e-basis}) reads
\begin{equation}\label{eq:gammaeps}
	\gamma_\mu \epsilon = \left( e^2 + \frac i2 \beta e^- \right)_\mu \epsilon
	+ \frac12 \left( e^+ - i \beta e^2 + \frac14 \beta^2 e^- \right)_\mu e^- \epsilon\ .
\end{equation}

Although it would be possible to proceed in this fashion, in what follows we found it easier to simply consider the null and timelike cases separately. This is actually a little simplistic, because the vector $z$ might very well be generically timelike, and null in some locus $N$ (we will indeed see one example of relevance to us where this happens). Strictly speaking, such cases are not covered by our analysis; in such cases, however, it is enough to work on $M_3-N$, and deduce formulas on $M_3$ by continuity.


\subsection{The null case} 
\label{sub:null}

We will first consider the case where $z$ is null, which, as we saw in section \ref{sub:alg}, corresponds to $\epsilon$ being a phase times a Majorana spinor. Using U(1) gauge transformations, we can actually reduce directly to the case where $\epsilon$ is Majorana. 
In this case, there will be no gauge field \footnote{As we saw in section \ref{sub:sv}, there is a slightly more general case where $\epsilon$ is a locally a phase times a Majorana spinor. In this case the gauge field is not zero
but just a flat connection.}:
\begin{equation}
	A=0\ .
\end{equation}

This case was actually already discussed in \cite{deMedeiros:2012sb} and we will be brief. We can simply use the strategy outlined at the end of subsection \ref{sub:alg}: introduce $e^-$ such that (\ref{eq:e-}), use the basis (\ref{eq:e-basis}) for spinors, and (\ref{eq:gammaeps}) to compute the action of $\gamma_\mu$. The latter is now much simpler because $\beta=0$. Notice in particular that
\begin{equation}
	e^+ = z\ .
\end{equation}
We can also define intrinsic torsions as in \cite[Sec.~3.2]{Cassani:2012ri}\footnote{We denote these $p$ and $q$ with a hat to distinguish them from those in the next subsection.}:
\begin{equation}
	\nabla_\mu \epsilon = \hat p_\mu \epsilon + \hat q_\mu e^- \epsilon\ .
\end{equation}
We find that the conformal Killing spinor equation (\ref{eq:cks}) reads
\begin{equation}\label{eq:ncks}
	\hat p_+= \hat q_-=0 \ ,\qquad \hat p_- = -2 \hat q_2 \ ,\qquad \hat p_2 = 2 \hat q_+ \ ,
\end{equation}
in a notation where any vector $V$ is expanded on the vielbein $\{ e^+, e^-, e^2\}$ as 
\begin{equation}\label{eq:index}
	V= V_+ z + V_- e^- + V_2 e^2\ .
\end{equation}
We can compute, on the other hand, 
\begin{equation}
	\nabla_\mu z_\nu = 2 (\hat p_\mu z_\nu -2 \hat q_\mu e^2_\nu )\ .
	\label{nullnablaz}
\end{equation}
Symmetrizing this and expanding all vectors as in (\ref{eq:index}), we get the equation for $z$ to be a conformal Killing vector, 
\begin{equation}
	({\cal L}_z g)_{\mu\nu}= 2 \nabla_{(\mu} z_{\nu)} = \lambda g_{\mu\nu}
\end{equation}
gives exactly the same equations as in (\ref{eq:ncks}) \cite{deMedeiros:2012sb}. This shows that an uncharged Majorana conformal Killing spinor exists if and only if there is a null conformal Killing vector.

This correspondence works without introducing any gauge field $A$. It is natural to wonder whether there is any way of extending the correspondence to the case where the gauge field $A$ is present, similarly to \cite{Cassani:2012ri}. The natural guess is that one should now allow $\epsilon$ to be any Dirac spinor; we now proceed to check this idea.\footnote{\label{foot:dirac}One can also consider the case where $A=0$ and the spinor $\epsilon$ is Dirac. This case is easier than the one we are about to consider in section \ref{sub:t}: if we write $\epsilon= \epsilon_1 + i \epsilon_2$, with $\epsilon_i$ Majorana, we see that the $\epsilon_i$ are separately CKS. Moreover, the vector $z$ associated to $\epsilon$ is $2 z_\mu = \overline{\epsilon} \gamma_\mu \epsilon = (\overline{\epsilon_1}- i \overline{\epsilon_2}) \gamma_\mu (\epsilon_1 + i \epsilon_2)= \overline{\epsilon_1} \gamma_\mu \epsilon_1 + \overline{\epsilon_2} \gamma \epsilon_2= z_1 + z_2$, where the $z_i$ are now null. So a CKV $z$ corresponds to a Dirac CKS with $A=0$ if and only if it can be written as the sum of two null CKVs.}


\subsection{The time-like case} 
\label{sub:t}

When $z$ is timelike, we observe that $\epsilon$ is not Majorana, not even up to a phase; this means that
\begin{equation}\label{eq:*basis}
	\epsilon \ ,\qquad \epsilon^*
\end{equation}
are independent spinors. We can thus use this basis, rather than (\ref{eq:e-basis}), to expand Dirac spinors. 

This also gives a rather natural choice for a vielbein: namely, one can define a complex vector $w$ via
\begin{equation}
	\epsilon^* \otimes \overline{\epsilon} = w \ ,\qquad w_\mu = \frac12 \overline{\epsilon} \gamma_\mu \epsilon^*\ .
\end{equation}
($\overline{\epsilon}\epsilon^*=0$, as can be seen in our basis by noting that $\gamma^0$ is antisymmetric.) Its action on $\epsilon$ can be worked out similarly to (\ref{eq:ze}): 
\begin{equation}
	w \epsilon = 2 i \beta \epsilon \ ,\qquad \bar w \epsilon  = 0\ .
\end{equation}
This also implies
\begin{equation}
	z \cdot w = w^2 = 0 \ ,\qquad \bar w \cdot w = 2 \beta^2 \ .
\end{equation}
Together with (\ref{eq:ze}), this means that $\frac1 \beta \{ z, {\rm Re} w, {\rm Im}  w \}$ is a vielbein. 

To proceed with our computation, we define new intrinsic torsions $p$ and $q$, this time via
\begin{equation}
	\nabla_\mu \epsilon = p_\mu \epsilon + q_\mu \epsilon^*\ .
\end{equation}
In terms of these, the  CKS equation (\ref{eq:cks}) becomes the following system: 
\begin{equation}\label{eq:pA}
	p^A_{\bar w} = q_w = 0 \ ,\qquad p^A_z = - q_{\bar w} \ ,\qquad
	p^A_w = - q_z \ ,
\end{equation}
where $p^A\equiv p-iA$. By taking linear combinations, we can eliminate $A$ from (\ref{eq:pA}) and obtain the following equations\footnote{When we write for example ${\rm Re} p_w$, we mean $({\rm Re} p)_w$ rather than ${\rm Re} (p_w)$.} on the geometry: 
\begin{equation}\label{eq:tcks}
	{\rm Re} (p_z + q_{\bar w}) = 0 \ ,\qquad 2 {\rm Re} p_w + q_z = 0 \ ,\qquad q_w = 0 \ .
\end{equation}
The remaining equations in (\ref{eq:pA}) determine $A$:
\begin{equation}\label{eq:At}
	A = {\rm Im} (p + q_{\bar w} z + q_z w)\ .
\end{equation}
It is easy to show that this gauge field is invariant under Weyl transformation of the metric. This is as it should be: the CKSs and CKVs are intrinsically defined for conformal geometry. 

Let us now compare the conditions (\ref{eq:tcks}) to a condition on $z$ alone. We have 
\begin{equation}
	\nabla_\mu z_\nu = 2 {\rm Re} ( p_\mu z_\nu + q_\mu w_\nu )\ .
\end{equation}
As in section \ref{sub:null}, symmetrizing this and expanding all vectors as in (\ref{eq:index}) we get the equation for $z$ to be a conformal Killing vector, 
\begin{equation}
	({\cal L}_z g)_{\mu\nu}= 2 \nabla_{(\mu} z_{\nu)} = \lambda g_{\mu\nu}\ ,
\end{equation}
gives exactly the same equations as in (\ref{eq:tcks}). 

We have thus proven that existence of a charged Dirac CKS on a three-dimensional manifold is equivalent to the existence of a timelike CKV. The CKS will be charged with respect to the gauge field $A$ in (\ref{eq:At}).

\subsection{Superalgebras} 
\label{sub:super}

We end this section with a brief discussion of the minimal superalgebras one gets in the null and timelike cases. 

Since in the null case we are discussing CKS solutions both in ${\cal N} = 1$ and ${\cal N} = 2$, the minimal number of independent components of the CKS can be just one. Therefore the smallest superconformal symmetry algebra for a SCFT on a manifold with a null CKV just consists of one bosonic and one fermionic generator, $B$ and $Q$:
\begin{equation}
[Q, B] = 0\ , \qquad \{Q, Q \} = B\ .
\end{equation}

In the timelike case, the CKS solutions are charged, and we are in the case of ${\cal N} = 2$ with a ${\rm U}(1)_R$-symmetry group. The CKS solution therefore consists of a minimum of two independent solutions that the R-symmetry acts upon. It follows that the minimal superalgebra for a SCFT on a manifold with a timelike CKV consists of the generator $H$ associated to the CKV, of the R-symmetry generator $R$, and of a complex supercharge ${\cal Q}$. Their nonvanishing commutation relations read
\begin{equation}
[{\cal Q}, R] = {\cal Q}\ , \qquad \{{\cal Q}, \bar{\cal Q} \} = H\ .
\end{equation}
This defines the superalgebra ${\rm U}(1|1)$.
 


\section{Introducing coordinates} 
\label{sec:coord}

In this section, we will make the general discussion more explicit by introducing coordinates, and we will write the general local form of the metric and of the gauge field of a background preserving supersymmetry. In the null case we recover the known classification of Lorentzian manifolds with a real conformal Killing spinor.

\subsection{Coordinates in the null case} 
\label{sub:coordnull}

The null case is characterized by the existence of a null conformal Killing vector $z$. With a Weyl rescaling we can always transform $z$ into a Killing vector.  We can introduce a natural coordinate $y$ such that as vector field 
\begin{eqnarray}
z & = & \frac{\partial}{\partial y}~.
\end{eqnarray}
We can further simplify our problem and introduce a second explicit coordinate. By antisymmetrizing (\ref{nullnablaz}), we
obtain
\begin{equation}
d z = 2 \hat p\wedge z - 4 \hat q \wedge e^2
\end{equation}
 and considering that $\hat q_-=0$ we find
 \begin{equation}
 z \wedge d z = 0\ .
 \end{equation}
This condition characterize  non-twisting geometries. By  Frobenius theorem, the distribution defined by vectors orthogonal to $z$ is integrable; hence there exist functions $F$ and $u$ such that
\begin{equation}
z =  F du \ .
\end{equation}
The metric then takes the form
\begin{equation} 
ds^2 = F du (dy + G du) + H dx^2\ ,
\end{equation}
where the functions $F,G,H$ depend only on $u$ and $x$.
With a further Weyl rescaling and a change of coordinates we can reduce the metric to the form
\begin{equation}
ds^2 = du ( dy + k(u,x) du) + dx^2 \, .
\end{equation}
We thus see that a metric supporting a Majorana conformal Killing vector is necessarily conformally equivalent to a pp-manifold. This reproduces the known classification \cite[Th.~5.1]{baum-leitner}.


\subsection{Coordinates in the time-like case} 
\label{sub:coordtime}

The time-like case is characterized by the existence of a time-like  conformal Killing vector $z$. As before, with a Weyl rescaling, we can transform $z$ into a Killing vector and choose a natural adapted coordinate 
\begin{eqnarray}
z & = & \frac{\partial}{\partial t}~.
\end{eqnarray}

If the norm of $z$ is not vanishing we can perform a further Weyl rescaling and set it equal to one. The most general
metric preserving supersymmetry is then locally conformally equivalent to
\begin{equation}\label{eq:gt}
ds^2 = -(d t +\omega )^2 + e^{2 \psi} ( d x^2 +d y^2 )
\end{equation}
where the function $\psi$ and the one-form $\omega$ only depend on $x$ and $y$. (We have used the fact that any two-dimensional metric is conformally flat, because of the uniformization theorem.) The corresponding vielbein is
\begin{eqnarray}
z & =& d t +\omega \\
w & =& e^{i\psi} ( dx + i dy )
\end{eqnarray}
Using (\ref{eq:At}) we can compute the gauge field
\begin{equation}\label{Atcoord}
A = -f (d t +\omega ) +\frac{1}{2}(\partial_x \psi dy - \partial_y\psi dx )
\end{equation}
and the corresponding curvature
\begin{equation}\label{eq:Ftcoord}
F = (d t +\omega ) df +\left ( \frac{1}{2}e^{-2 \psi} \Delta \psi - 2 f^2 \right ) {\rm vol}_2\ ,
\end{equation}
where $ {\rm vol}_2= e^{2\psi} dx \wedge dy$ is the volume of two-dimensional base and the function $f$ is defined by
\begin{equation}\label{eq:dof}
d \omega = 2 f  \, {\rm vol}_2\ .
\end{equation}

The previous analysis applies to any local patch where the norm of $z$ is not vanishing. To describe the general case we can introduce 
a $t$-independent Weyl factor, so that 
\begin{equation}\label{eq:gtc}
ds^2 =  g^2 \left (-(d t +\omega )^2 + e^{2 \psi} ( d x^2 +d y^2 )\right )\ ,
\end{equation}
where singularities in $g$ correspond to the zeros of $z$. The gauge field, being invariant under Weyl transformation, is still given by
equation (\ref{Atcoord}). In section \ref{sub:rs2class} we will see an example of a CKV for which it is necessary to use ({\ref{eq:gtc}).

\subsection{From bulk to boundary}
\label{sec: bulktoboundary}

It is of some interest in view of holographic applications to compare our boundary results with the known classifications of BPS solutions of minimal ${\cal N}=2$ gauged supergravity in four dimensions \cite{Caldarelli:2003pb,Cacciatori:2004rt}.

We consider solutions of four-dimensional gauged supergravity with a non trivial profile for the metric and the graviphoton $A$
with asymptotic behavior as in equations (\ref{asymp}), (\ref{asympA}). These solutions are the dual description of a CFT on  $M_3$
with a non vanishing background value for the R-symmetry gauge field. As recalled in section \ref{sub:qft}, the bulk supersymmetry conditions  reduce on the boundary  to the equation for the existence of a charged conformal Killing spinor \cite{Klare:2012gn}. 
We conclude that the boundary metric of a supersymmetric solution of gauged supergravity with asymptotics (\ref{asymp}), (\ref{asympA}) must be of the form discussed in sections \ref{sub:coordnull} and \ref{sub:coordtime}. 
It is interesting to see explicitly how  it works for the general   BPS solutions discussed in \cite{Caldarelli:2003pb,Cacciatori:2004rt}.

The supersymmetric bulk solutions  found in \cite{Caldarelli:2003pb,Cacciatori:2004rt} fall into two different classes, characterized by the existence of a null or time-like Killing vector, which reflect the analogous dichotomy on the boundary.    The general supersymmetric solution in the null class can be written as \cite{Cacciatori:2004rt}
\begin{equation}
	\begin{split}
		ds_4^2 &= \frac{1}{z^2} \left [ {\cal F}(u,z,x) du^2 + 2 du dx + dz^2 + dx^2 \right ]\ , \\
		A &=  \Phi (u) dz\ ,
	\end{split}
\end{equation}
with
\begin{equation}
\label{constrK}
\partial_z^2 {\cal F} +\partial_x^2 {\cal F}  - \frac{2}{z} \partial_z {\cal F} = - 4 z^2 \left (\frac{\partial \Phi}{\partial u} \right )^2 \, .
\end{equation}
We obtain asymptotically AdS background by choosing $r=1/z$ and ${\cal F}= f(x,u) + O(z^2)$. The gauge field vanishes
at the boundary $A_\mu =0, A_r = O(1/r^2)$  and the metric becomes asymptotically
\begin{equation}
ds_4^2  \sim     r^2 \left [ f(x,u) du^2 + 2 du dx + dx^2  \right ] +\frac{dr^2}{r^2}
\end{equation}
showing that the boundary metric is a pp-manifold. This is in agreement with our general classification in section  \ref{sub:coordnull}.
The differential equation (\ref{constrK}) fix order by order the  subleading terms in the expansion of the metric and the gauge field. 

The general supersymmetric solution in the time-like class can be written as \cite{Cacciatori:2004rt} \footnote{We add an index $B$ in order to distinguish the bulk quantities ($B$) from the boundary ones.} 
\begin{equation}
	\begin{split}
		ds_4^2 &= -\frac{4}{|F_B|^2} ( d t +   \omega_B )^2 +\frac{|F_B|^2}{4} \left ( dr^2 + e^{2  \phi_B} ( dx^2 + dy^2 ) \right ) \\
		F_B &=  ( d t +  \omega_B )\wedge d  f_B  +  *  \left (  ( d t +  \omega_B )\wedge (d  g_B + d r )\right )
	\end{split}
\end{equation}
where $F_B=-2/(g_B+ i f_B)$. The functions $F_B$, $\phi_B$ and the one-form $\omega_B$ satisfy a set of differential equations (see   \cite[Eqs.(2.2)--(2.5)]{Cacciatori:2004rt}).
We obtain asymptotically AdS background by choosing
\begin{equation}
	\begin{split}
		g_B= - r + g(x) + \cdots \ ,\qquad    
		f_B =  f(x) + \cdots \ , \\
		e^{2 \psi_B} = r^4 e^{2 \psi(x)} +\cdots \ ,\qquad  \omega_B = \omega(x) + \cdots\ .
	\end{split}
\end{equation}
Using \cite[Eqs.(2.2)--(2.5)]{Cacciatori:2004rt} we can check that  the asymptotic form of the metric and the gauge fields are
\begin{equation}
	\begin{split}
		ds_4^2 &\sim r^2 \left [-(d t +\omega )^2 + e^{2 \psi} ( d x^2 +d y^2 ) \right ] +\frac{dr^2}{r^2} \\
		F &\sim (d t +\omega ) df +\left ( \frac{1}{2}e^{-2 \psi} \Delta \psi - 2 f^2 \right ) {\rm vol}_2
	\end{split}
\end{equation}
with $ d \omega = 2 f  \, {\rm vol}_2$, in complete agreement with equations (\ref{eq:gt}) and (\ref{eq:Ftcoord}). \cite[Eqs.(2.2)--(2.5)]{Cacciatori:2004rt}
fix order by order the other subleading functions in the expansion of the metric and the gauge field.



\section{Example: conformally flat spacetimes} 
\label{sec:examples}

As an example of our general formalism, we will now consider some conformally flat spacetimes: $\rr \times S^2$, $\rr^{1,2}$ and $\rr \times H_2$ (where $H_2$ is hyperbolic two-space). For these spaces the group of conformal isometries is ${\rm SO}(3,2)$ and we can apply our results explicitly since expressions for its generators are already known. These geometries are very simple, but already here we will see that there are several different ways to write superconformal theories, some of which unexpected. 

We will actually mostly focus on $\rr\times S^2$, both because of its applications to black hole physics (see section \ref{sec:bh}) and because $\rr^{1,2}$ and $\rr\times H_2$ are conformal to patches of it. 

We will begin by giving in section \ref{sub:rs2mon} and \ref{sub:rs20} two simple examples of timelike and null conformal Killing vectors, and of their associated conformal Killing spinors. The example in section \ref{sub:rs2mon} corresponds to a constant field-strength on $S^2$, while the examples in \ref{sub:rs20} have vanishing gauge field. In section \ref{sub:rs2class} we will derive a classification of CKSs on $\rr\times S^2$. In particular we will show that the two simple classes discussed in sections \ref{sub:rs2mon} and \ref{sub:rs20} are the only possibilities which preserve the full rotational symmetry SO(3) of the sphere $S^2$. As anticipated in the introduction, they will correspond holographically to static black holes, with a gauge field which is either
\begin{equation}
	F= - \frac12 {\rm vol}_2\qquad {\rm or} \qquad F=0\ .
\end{equation}
We will actually see that there is also a family that interpolates between these two cases. This family corresponds holographically to rotating black holes.
We conclude by discussing more briefly some examples of CKSs for Minkowski$_3$ in section \ref{sub:r3} and for $\rr\times H_2$ in section \ref{sub:rh2}.

\subsection{Monopole gauge field on $\rr\times S^2$} 
\label{sub:rs2mon}

The metric on $\rr \times S^2$ is
\begin{equation}\label{eq:rs2}
	-d\tau^2 + d s^2_{S^2}=-d\tau^2 + d\theta^2 + \sin^2(\theta) d\phi^2\ .
\end{equation}

As we mentioned above, the group of conformal isometries of $\rr\times S^2$ is ${\rm SO}(3,2)$; each generator of its Lie algebra ${\rm so}(3,2)$ is a CKV. The timelike condition restricts us to an open set in this Lie algebra; so there is a ten-dimensional space of timelike CKVs $z$. As we saw in section \ref{sub:t}, each timelike CKV will give a CKS charged with respect to a certain gauge field. One way to compute this gauge field is to change coordinates so that the timelike CKV is expressed as $\del_t$, as we did in section \ref{sub:coordtime}. 

We begin in this subsection with the easiest timelike CKV of (\ref{eq:rs2}), namely the one which reads
\begin{equation}\label{eq:deltau}
	z=\del_\tau
\end{equation}
already in the original coordinates. This corresponds, in the general form (\ref{eq:gtc}), to taking $t=\tau$, and
\begin{equation}\label{eq:psiS2}
	g=1 \ ,\qquad \omega= 0 \ ,\qquad e^{2 \psi} = \frac4{(1 + x^2 + y^2)^2}\ ;
\end{equation}
the metric $ds^2_{S^2}$ is then expressed as the Fubini-Study metric on $\cc\pp^1$. As we will see, other timelike vectors will correspond to writing the original metric (\ref{eq:rs2}) in different coordinates, with more complicated choices of $g, \omega$ and $\psi$. This is not always the most practical way of computing the gauge field, but it will suffice for this subsection and the next.

In the present case, (\ref{eq:psiS2}) gives $f=0$; from the general expression (\ref{eq:Ftcoord}), we get 
\begin{equation}
	F= -\frac12 {\rm vol}_2 \ .
\end{equation}
This is simply the monopole field-strength on $S^2$. So the CKV (\ref{eq:deltau}) will give rise to a CKS $\epsilon$ charged with respect to the connection $A= \frac12\cos(\theta) d \phi$. In other words, $\epsilon$ will be a section not simply of the spinor bundle $\Sigma$, but of $\Sigma \otimes {\cal O}(1)$. This bundle is trivial. In field theory, this is usually referred to as a ``twist''; in the context of supergravity solutions, it is referred to as ``the gauge field canceling the spin connection''. In the CKS equation (\ref{eq:cks}), this manifests itself as the fact that $\nabla^A_\mu$ reads $\nabla^A_\theta = \del_\theta$, $\nabla^A_\phi = \del_\phi - \frac12 \cos(\theta) (\gamma^0 + i)$. One can then solve (\ref{eq:cks}) simply by taking $\epsilon$ to be constant and annihilated by $(\gamma^0 + i)$:
\begin{equation}\label{monopole-cks}
	\epsilon= (1 + i \gamma^0) \epsilon_0\ .
\end{equation}
This describes a solution with two independent CKS components. The resulting SCFT has a ${\rm U}(1|1)$ symmetry algebra, as discussed in the end of section \ref{sub:coordtime}. Due to the form of the field strength, the rotational symmetry in the spatial part of the metric is preserved. The rotations however act trivially on the CKS components, as can be easily seen from \eqref{monopole-cks}. The full superconformal symmetry group is therefore ${\rm U}(1|1) \times {\rm SO}(3)$.


\subsection{Zero gauge field} 
\label{sub:rs20}

A second easy example we can consider is when the gauge field vanishes, $A=0$. 

In this case, we can consider CKSs that are Majorana; as we saw in section \ref{sub:null}, these correspond to CKVs which are null. We can also consider CKSs which are Dirac; as we mentioned in footnote \ref{foot:dirac}, these correspond to timelike CKVs which can be written as a sum of two null CKVs. 

In section \ref{sub:rs2class}, we will see how to obtain CKVs of both types systematically. However, on $\rr\times S^2$, uncharged CKSs can be obtained also directly by solving the original equation (\ref{eq:cks}). The idea is to start from the Killing spinors on $S^2$, and to add a suitable dependence on time. One finds two independent CKS solutions with a total of 8 independent components:
\begin{equation}\label{eq:CKSrs2}
	\epsilon_{\mathbb{R} \times S^2} = 
							(1+\gamma_0) e^{i t/2} \epsilon_1 +
							(1-\gamma_0) e^{- i t/2} \epsilon_2\ .
\end{equation} 
where $\epsilon_i$ are two Killing spinors on $S^2$. An explicit expression for them can be found for example in \cite{lu-pope-rahmfeld}:
\begin{equation}
	\epsilon_i = e^{i \gamma_1 \theta/2} e^{-\gamma_0 \varphi/2} \epsilon_i^0\ ,
\end{equation}
where $\epsilon_i^0$ are two arbitrary constant Dirac spinors.

The CKVs that correspond to these CKSs via (\ref{eq:betaz}) in general are a bit complicated, but we will see them emerge in the general classification in section \ref{sub:rs2class}. Some examples are particularly easy: for example, if $\epsilon_2=0$, one gets a vector of the form $\del_\tau + J$, where $J$ is a Killing vector of the sphere (a rotation around some axis). If both $\epsilon_1$ and $\epsilon_2$ are non-zero, the cross-terms between them in the bilinear $z_\mu$ generate linear combinations of the conformal Killing vectors of the sphere, whose generators $K_i$ are introduced below in (\ref{eq:Ki}). This is in accordance with the fact that this background preserves the maximal number of CKVs and CKSs possible in three dimensions: the conformal symmetry group is ${\rm SO}(3,2)$, while the conformal supergroup is ${\rm OSp}(2|4)$ (for ${\cal N} = 2$).


\subsection{Classification of null and time-like CKVs on $\rr\times S^2$} 
\label{sub:rs2class}

We will now look more systematically at the null and timelike CKVs on $\rr\times S^2$. We will see that, in a certain sense, the only possibility is a one-parameter family that contains the cases discussed in sections \ref{sub:rs2mon} and \ref{sub:rs20}. 

The reason we can perform explicit computations on $\rr\times S^2$ is that the CKVs are known. They can be obtained for example from the Killing vectors
\begin{equation}
	v_{IJ}\ ,\qquad I,J=1,\ldots,5
\end{equation}
on AdS$_4$. These vectors generate the Lie algebra ${\rm so}(3,2)$; a common convention, which we will follow, is to take the quadratic form to be negative in directions $1$ and $5$, and positive in directions $2,3,4$. The explicit expression of the $v_{IJ}$ can be found for example in \cite{henneaux-teitelboim}. By taking a limit to the boundary, one can also find explicit expressions for the $v_{IJ}$ on $\rr\times S^2$. 
\begin{equation}\label{eq:vIJ}
	\begin{split}
		& v_{51}= \del_\tau \ ,\qquad v_{i1}= \cos(\tau) Y_i \del_\tau + \sin(\tau) K_j  \ ,\\
		& v_{ij}= -\epsilon_{ijk} J_k \ ,\qquad 
		 v_{i5}= -\sin(\tau) Y_i \del_\tau + \cos(\tau) K_j\ ,
	\end{split}
\end{equation}
where indices $i,j,k$ range over $2,3,4$; $J_i$ are the Killing vectors of $S^2$ that correspond to rotating it around the three axes; $Y_i$ are the three spherical harmonics with $\ell=1$, and $K_i$ are three conformal Killing vectors of $S^2$, which are generated by the $Y_i$: 
\begin{equation}\label{eq:Ki}
	K_i = -\frac12 g_{S^2}^{-1} d Y_i\ .
\end{equation}
The $J_i$ and $K_i$ together generate the group ${\rm Sl}(2,\rr)$ of M\"obius transformations of $S^2$. 

We should now consider a general conformal Killing vector
\begin{equation}
	z= \sum_{I,J} c_{IJ} v_{IJ}\ ,
\end{equation}
impose that it is never spacelike on $\rr\times S^2$, and then compute the gauge field $A$ according to the procedure outlined in section \ref{sub:t} if it is generically timelike, or simply conclude that $A=0$ if it is everywhere null. 

The computation looks very complicated. Fortunately, we can use the symmetries of the problem to simplify it considerably. Of course, for example, if two vectors are related by a rotation of $S^2$, the corresponding gauge fields will also be related by a rotation, and we can essentially compute the gauge field for only one of them. We can use in a similar way also conformal transformations. Let us formalize this as follows. The data to compute the gauge field-strength $F$ are the metric $g$ and a CKV $z$; to emphasize this dependence, we can write $F_{g,z}$. Suppose we have two CKVs $z$ and $z'$ of the same metric $g$ which are related by a conformal transformation, namely by an element of the group of conformal isometries $C \in {\rm SO}(3,2)$: 
\begin{equation}
	z' = C z\ .
\end{equation}
$C$ will in general take $g$ to $e^f g$ for some function $f$. So we have that $F_{g, z}= C F_{e^f g,z'}$. As we remarked in section \ref{sub:t}, however, $F$ is invariant under Weyl rescalings of the metric: so $F_{e^f g,z'}= F_{g,z'}$, and we have obtained 
\begin{equation}
	F_{g,z}= C F_{g,z'}\ .
\end{equation}
Summing up, if two CKVs $z$ and $z'$ are related by a conformal transformation $C$, their associated gauge field-strengths are also related by $C$. This suggests that we can just work up to the action of the conformal group ${\rm SO}(3,2)$ on the space of CKVs. The latter is just a copy of the Lie algebra ${\rm so}(3,2)$, so in fact this action is just the adjoint action of a group on its associated Lie algebra. 

All this means that we can limit ourselves to considering a $z$ for every orbit of the action of ${\rm SO}(3,2)$ on its Lie algebra ${\rm so}(3,2)$. So we need a classification of these orbits. Were we considering ${\rm SO}(5)$ instead of ${\rm SO}(3,2)$, the answer would be clear: the Lie algebra ${\rm so}(5)$ consists of antisymmetric matrices, and any antisymmetric matrix can be put in a block-diagonal form ${\rm diag}(a J_2, b J_2, 0)$ (where $J_2 = \left(\begin{smallmatrix}0 & -1 \\ 1 & 0 \end{smallmatrix}\right)$) by the adjoint action of a suitable orthogonal matrix. For orthogonal groups of mixed signature such as ${\rm SO}(3,2)$, the answer is less widely known, but fortunately still available in the literature: it is reviewed in \cite{figueroaofarrill-simon}, based on earlier literature, especially \cite{boubel} (see also \cite{djokovic-patera-winternitz-zassenhaus}). 

In particular, the classification of adjoints orbits for ${\rm SO}(3,2)$ in \cite[Sec.~4.2.1]{figueroaofarrill-simon} consists of a list of 15 cases, some of which are one-dimensional families (some are limiting cases of others: keeping account of this, the list could be shortened to 10 cases). We will not repeat the entire list here, because several of its items are vectors which are spacelike somewhere on $\rr\times S^2$, and hence cannot be used to produce CKSs on $\rr\times S^2$.\footnote{As we will review later, $\rr^{1,2}$ and $\rr \times H_2$ are conformal to certain patches in $\rr\times S^2$. Some of the cases that we are discarding here are spacelike outside of those regions, and can hence be used to produce CKSs in those spaces. Although we will not give a complete classification for those spaces, we will see an example of this in section \ref{sub:rh2}.} This actually leaves only three cases. 
\begin{enumerate}
	\item The first case corresponds to item (3) in \cite[Sec.~4.2.1]{figueroaofarrill-simon}. After adapting their notation to ours, and after performing a rotation on $S^2$ for convenience, it is 
	\begin{equation}\label{eq:z3r}
		z=v_{51} - v_{45} = (1- \cos(\tau) \cos(\theta))\del_\tau + \sin(\tau)\sin(\theta) \del_\theta \  
	\end{equation}
	where we have recalled the definition of the $v_{IJ}$ in (\ref{eq:vIJ}). In the patch of $\rr\times S^2$ which is conformal to Minkowski$_3$, this vector actually becomes simply $\del_t$. In flat space, our general expression (\ref{eq:At}) for the gauge field obviously gives $A=0$. So this $z$ corresponds to a CKS with $F=0$, which is actually the one in (\ref{eq:CKSrs2}) with $\epsilon^0_1= \frac1{2\sqrt{2}}e^{ia}{- i \choose 1}$, $\epsilon^0_2 = -\frac1{2\sqrt{2}}e^{ia}{1 \choose i}$, for any real $a$. 

	\item The second case corresponds to item (8) in \cite[Sec.~4.2.1]{figueroaofarrill-simon}. In our notation:
	\begin{align}\label{eq:z8}
		z&= (1+\alpha) v_{51} + (1-\alpha) v_{23} - v_{12} + v_{53}\qquad \qquad (\alpha\ge 0)\\
		&=(1+\alpha - \cos(\tau-\phi)) \del_\tau - \cos(\theta) \sin(\tau - \phi) \del_\theta
		+ \left(\alpha - 1 + \frac{\cos(\tau-\phi)}{\sin(\theta)}\right)\del_\phi\ . \nonumber
	\end{align}
	
	For $\alpha=0$, this $z$ is lightlike everywhere, and again corresponds to a CKS with $F=0$, which is the one in (\ref{eq:CKSrs2}) with $\epsilon^0_1= \frac1{2\sqrt{2}}e^{ia}{1 \choose i}$, $\epsilon^0_2 = -\frac1{2\sqrt{2}}e^{ia}{i \choose 1}$. For $a=0$, the CKS is Majorana. 
	
	For $\alpha>0$, $z$ in (\ref{eq:z8}) is generically timelike, and lightlike at $\{ \theta=\frac\pi2, \phi=\tau \}$. The corresponding $F$ is rather hard to compute explicitly, but it diverges at the lightlike locus. 
	
	\item The third case corresponds to item (10) in \cite[Sec.~4.2.1]{figueroaofarrill-simon}, and it is simply
	\begin{equation}
		z= v_{51}+ \alpha v_{32} = \del_\tau + \alpha \del_\phi \  \qquad (0\le\alpha \le 1)\ .
	\end{equation}
	For $0\le\alpha < 1$, this CKV is timelike everywhere. For $\alpha=1$, it becomes null at the equator $\{ \theta=\frac\pi 2\}$. In one gauge, the gauge field we compute from our general expression (\ref{eq:At}) for $A$ is 
	\begin{equation}
		A= \frac{\cos(\theta)}{2\sqrt{1-\alpha^2 \sin^2(\theta)}} (\alpha d \tau + d\phi)\ ,
	\end{equation}
whose field-strength is 
	\begin{equation}\label{eq:F10}
		F= \frac{(\alpha^2-1)}{2(1-\alpha^2 \sin^2(\theta))^{3/2}} \sin(\theta) d\theta \wedge ( \alpha d\tau + d \phi)\ .
	\end{equation}
	
	There are two notable cases in this family: the extreme values, $\alpha=0$ and $\alpha=1$. 
	\begin{itemize}
		\item For $\alpha=0$, we have $\del_\tau$, which we have already analyzed in section \ref{sub:rs2mon}: it corresponds to a gauge field-strength $F=-\frac12\vol_2$. 

		\item For $\alpha=1$, as we mentioned $z_{\alpha=1}=\del_\tau + \del_\phi$ becomes lightlike at the equator $\{ \theta=\frac\pi 2\}$. As we can see from (\ref{eq:F10}), the gauge field-strength vanishes. In fact, this could be foreseen by remembering footnote \ref{foot:dirac} and writing $z_{\alpha=1}$ as the sum of two null vectors: $v_{51}+v_{23}-v_{12}+v_{53}$ (which appeared in (\ref{eq:z8}) for $\alpha=0$) and its friend $v_{51}+v_{23}+v_{12}-v_{53}$. 
		
		Since in this case $F$ vanishes, the corresponding CKS is uncharged; and, as we anticipated at the end of section \ref{sub:rs20}, it is a particular case of (\ref{eq:CKSrs2}), this time for $\epsilon^1_0=\frac12 e^{ia} {i \choose 0}$, $\epsilon^2_0=0$. 
		
		Since the norm of $z$ goes to zero at the equator, this case also provides an example where it is necessary to use (\ref{eq:gtc}) rather than the simpler (\ref{eq:gt}). It can be checked that the metric (\ref{eq:rs2}) of $\rr\times S^2$ can be written in the form (\ref{eq:gtc}), with
		\begin{equation}
			g=1-|\zeta|^2\ ,\qquad
			\omega= \frac{|\zeta|^2}{1-|\zeta|^2} d \phi \ ,\qquad
			e^{2 \psi}= \frac1{(1-|\zeta|^2)^2}\ ,
		\end{equation}
		where $\zeta=\sin(\theta)e^{i \phi}$.
	\end{itemize}
\end{enumerate}

This classification shows that the only cases for which the gauge-field strengths are invariant under the rotation group SO(3) of the sphere $S^2$ are the case $F=0$ (which we had already considered in section \ref{sub:rs20}) and the case $F=-\frac12{\rm vol}_2$ (which we had already considered in section \ref{sub:rs2mon}). If we do not impose the condition of SO(3) invariance, we also have a family for which $F$ is given by (\ref{eq:F10}), and a family for which $F$ is singular on a codimension-two locus. 
	 

\subsection{Minkowski$_3$} 
\label{sub:r3}

We will now consider more briefly the case of Minkowski$_3$ space, namely $\rr^{1,2}$ with metric
\begin{equation}\label{eq:flat}
	ds^2= - dt^2 + dx^2 + dy^2\ .
\end{equation}
As is well-known, this is conformal to the region
\begin{equation}\label{eq:minkdiamond}
	|\theta-\pi|+|\tau|<\pi    
\end{equation}
of $\rr\times S^2$, which we can also write alternatively as the region where $\cos(\tau)\ge \cos(\theta)$. This region starts existing at $\tau=-\pi$, when it consists of the south pole $\{\theta=\pi\}$, expands from there until it consists of the whole $S^2$ at $\tau=0$, and contracts until it shrinks to the south pole again at $\tau=\pi$. If one draws $\rr\times S^2$ as a cylinder, the region (\ref{eq:minkdiamond}) is often depicted as a diamond drawn on that cylinder. This diamond is the Penrose diagram of flat Minkowski space. The change of coordinates 
\begin{equation}\label{eq:ms}
	t= \frac{\sin(\tau)}{\cos(\tau)-\cos(\theta)} \ ,\qquad
	x= \frac{\sin(\theta)}{\cos(\tau)-\cos(\theta)} \sin(\phi) \ ,\qquad
	y= \frac{\sin(\theta)}{\cos(\tau)-\cos(\theta)} \cos(\phi)
\end{equation}
takes the flat Minkowski metric (\ref{eq:flat}) to 
\begin{equation}
	ds^2= \frac1{(\cos(\tau)-\cos(\theta))^2} (-d\tau^2 + d\theta^2 + \sin^2(\theta) d \phi^2)\ ,
\end{equation}
which is related by a Weyl transformation to the metric (\ref{eq:rs2}) of $\rr \times S^2$. (For $\tau=0$, this is a stereographic projection from $S^2$ to $\rr^2$.) 
The change of coordinates (\ref{eq:ms}) is also often expressed in terms of the polar coordinate $\rho \equiv \sqrt{x^2 + y^2}=\frac{\sin(\theta)}{\cos(\tau)-\cos(\theta)}$, and of light-cone coordinates:
\begin{equation}
	t+ \rho = \cot \left(\frac{\theta - \tau}2\right)
	\ ,\qquad
	t- \rho = \cot \left(\frac{\theta + \tau}2\right)\ .
\end{equation}
For more details, see for example \cite[Sec.~5.1]{hawking-ellis},\cite[Sec.~2.2]{magoo}.

The group of conformal isometries is again ${\rm SO}(3,2)$; its generators can still be taken to be the $v_{IJ}$ in (\ref{eq:vIJ}), in the new coordinates. Their interpretation is now completely different: taking $i=5,2,3$, and $x^5=t$, $x^2=x$, $x^3=y$, we have that $v_{i4}+v_{i1}$ are the momentum generators; $v_{i4}-v_{i1}$ generate the special conformal transformations; $v_{ij}$ generate the Lorentz group; and $v_{14}$ is the dilatation operator. The finite action of the group is actually not well-defined on Minkowski$_3$ space itself, but rather on an appropriate conformal compactification. An example of this is given by a finite special conformal transformation $x \to \frac{x + b x^2}{1+2 b\cdot x + b^2 x^2}$, which diverges at a certain surface (the locus where the denominator goes to zero). Another way of realizing this is to think of the diamond (\ref{eq:minkdiamond}) in $\rr\times S^2$; most finite transformations will not keep it invariant. For example, a rotation that moves the ``north pole'' $\{\theta=0\}$ to some other point will not leave the diamond invariant. Some points which were in the interior of the diamond will be taken to its boundary: on these points the finite action in Minkowski$_3$ will diverge. Some other points will be taken out of the original diamond. But one can tile $\rr\times S^2$ with other diamonds, which can be identified with the original one (\ref{eq:minkdiamond}); this identification then gives the finite action of the group ${\rm SO}(3,2)$. 

All this means that we should be more cautious in repeating the argument in section \ref{sub:rs2class}. Certainly the three cases we discussed in that section, which are everywhere timelike or null in $\rr\times S^2$, can be mapped to Minkowski$_3$. There might be, however, other cases that we are missing in this way, because there might be CKVs which are timelike or null in the diamond (\ref{eq:minkdiamond}), and which we discarded in section \ref{sub:rs2class} because they were spacelike somewhere else. Moreover, even within the same adjoint orbit, one CKV might be timelike in the original diamond (\ref{eq:minkdiamond}), and another might be timelike in some other original diamond. So the choice of representatives done in \cite[Sec.~4.2.1]{figueroaofarrill-simon} might lead one to miss some possibilities. We will see an example of both these phenomena in section \ref{sub:rh2}. 

For these reasons, we have not attempted a classification for $\rr^{1,2}$. Let us instead comment about a couple of examples, both obtained from the cases we found in section \ref{sub:rs2class}.

The first example is $z=\del_t$. In this case, as we noticed already in section \ref{sub:rs2class}, our general expression for $A$ gives trivially zero, since everything is constant. We encountered this case already in section \ref{sub:rs2class}, as case 1 in our list there. The fact that there are uncharged CKSs in flat space is of course not surprising; they are the usual supercharges of a superconformal theory in flat space with a superconformal symmetry algebra ${\rm OSp}({\cal N}|4)$. Explicitly, for our case of ${\cal N} = 2$, the CKSs in the flat coordinates \eqref{eq:flat} are
\begin{equation}
	\varepsilon = \varepsilon'_0 - (t \gamma_0 + x \gamma_1 + y \gamma_2) \varepsilon''_0\ , 
\end{equation}
with the eight fermionic degrees of freedom packaged in two arbitrary constant spinors $\varepsilon'_0, \varepsilon''_0$.
	
The second example we want to comment on is more surprising. We can consider the vector $z=v_{51}$, which was equal to $\del_\tau$ on $\rr\times S^2$, and map it to Minkowski$_3$ using (\ref{eq:ms}). The result can be expressed in polar coordinates $\rho=\sqrt{x^2+y^2}$, $\phi=\arctan(x/y)$: 
\begin{equation}
	F= \frac{4\rho (-2t \rho dt + (1 +t^2 + \rho^2) d\rho)\wedge d\phi}{(1+(t^2-\rho^2)^2 +2(t^2+\rho^2))^{3/2}}\ .
\end{equation}
Not only is this $F$ non-zero, it also retains memory of it being topologically non-trivial on $\rr\times S^2$. For $t=0$, for example, we see that $F=\frac{4\rho}{(1+\rho^2)^2} d\rho \wedge d\phi$ is the volume form of the Fubini--Study metric on $\rr^2$ thought of as $\cc\pp^1$ minus the point at infinity. So it integrates to $2\pi$ on the $\rr^2$ at $t=0$, and hence on every spatial slice. As we already saw in section \ref{sub:rs2mon}, SCFTs on such a background exhibit ${\rm U}(1|1) \times {\rm SO}(3)$ symmetries, even though this fact is much less intuitive to see in Minkowski$_3$.

So we see that even on flat Minkowski$_3$ space there exist non-zero gauge fields to which one can couple a superconformal theory, such that it retains some supercharges.


\subsection{$\rr\times H_2$} 
\label{sub:rh2}

We will finally consider $\rr\times H_2$, with coordinates $\{T,\chi,\phi\}$ and metric
\begin{equation}\label{eq:rh2metric}
	ds^2= -dT^2 + d\chi^2 + \sinh^2(\chi) d\phi^2 \ .
\end{equation}
The change of coordinates from Minkowski$_3$ to $\rr\times H_2$ reads
\begin{equation}\label{eq:hm}
	t=e^T \cosh(\chi) \ ,\qquad
	x=e^T \sinh(\chi) \sin(\phi)\ ,\qquad
	y=e^T \sinh(\chi) \cos(\phi)\ .
\end{equation}
This indeed takes the flat space metric (\ref{eq:flat}) to $e^{2T} ds^2$, where $ds^2$ is given by (\ref{eq:rh2metric}). If we invert (\ref{eq:hm}), we get $T=\frac12 \log(t^2-x^2-y^2)$, $\chi={\rm arccoth}(t/\sqrt{x^2 + y^2})$, $\phi=\arctan(x/y)$. We see that this coordinate change only makes sense above the light cone, where $t>\sqrt{x^2+y^2}$. This is a sub-diamond of (\ref{eq:minkdiamond}). 

The generators of SO(3,2) now act as follows. Defining $\mu=2,3,5$: 
\begin{equation}\label{eq:vIJh}
	\begin{split}
			&v_{41}=\del_T \ ,\qquad v_{1\mu}=\cosh(T) Y_\mu \del_T -\sinh(T) K_\mu\ ,\\
			&v_{\mu\nu}= \epsilon_{\mu\nu}{}^\rho J_\rho \ ,\qquad
			v_{4\mu}= -\sinh(T) Y_\mu \del_T +\cosh(T) K_\mu\ .
	\end{split}
\end{equation}
where $J_\rho$ are the three generators of the ${\rm SO}(2,1)\cong {\rm Sl}(2,\rr)$ symmetry of $H_2$, $Y_\mu$ are the harmonic functions $\{ \sinh(\chi)\sin(\phi), \sinh(\chi) \cos(\phi), \cosh(\chi)\}$, 
and $K_\mu$ are three conformal Killing vectors of $H_2$, generated by the $Y_\mu$: 
\begin{equation}\label{notimportant}
	g_{H_2}^{-1} d Y_\mu = - K_\mu\ .
\end{equation}
(Notice the similarity of these formulas with (\ref{eq:vIJ}); this structure is a general way to promote CKVs of a manifold $M$ to CKVs of $\rr\times M$.)

In trying to give a classification of possible CKVs, the same difficulties arise as the ones we described for Minkowski$_3$ in section \ref{sub:r3}. Actually, we can also give an explicit example of some of the phenomena described there. The time translation vector
\begin{equation}
	\del_T=v_{41}
\end{equation}
is not in any of the classes described in section \ref{sub:rs2class}. That is because, although it is of course timelike in $\rr\times H_2$, it is not timelike on the whole of $\rr\times S^2$, but only in its small patch that ends up being conformal to $\rr\times H_2$. In the classification of \cite[Sec.~4.2.1]{figueroaofarrill-simon}, this would be item (2). The full symmetry group in this case is ${\rm U}(1|1) \times {\rm SO}(2,1)$ and, as already pointed out, corresponds to a different solution that we did not encounter in section \ref{sub:rs2class}.  

This example actually illustrates another subtlety we described in section \ref{sub:r3}. The representative chosen in \cite[Sec.~4.2.1]{figueroaofarrill-simon} for item (2) is actually, in our language, $v_{12}$. With the expression (\ref{eq:vIJh}), we can see that this CKV is actually spacelike at $\chi=0$. On $\rr\times S^2$, it would be timelike in some other patch conformal to $\rr\times H_2$, not in the one we have chosen, unlike $v_{41}$. 

For all these reasons, we have not attempted to give a classification for $\rr\times H_2$, either. As for Minkowski$_3$, of course, all three cases described in section \ref{sub:rs2class} can be mapped to $\rr\times H_2$, and will remain timelike or null. 

As for the particular vector $\del_T$, however, it is easy to see to which gauge field it corresponds; the discussion is very similar to $\del_\tau$ in the $\rr\times S^2$ case. The metric of $\rr\times H_2$ in (\ref{eq:rh2metric}) can be cast in the general form (\ref{eq:gtc}), to taking $t=T$, and
\begin{equation}\label{eq:psih2}
	g=1 \ ,\qquad \omega= 0 \ ,\qquad e^{2 \psi} = \frac4{(1 - (x^2 + y^2))^2}\ ;
\end{equation}
the metric $ds^2_{H_2}$ is then expressed as the Poincar\'e metric on the unit disc. The gauge field can then be obtained from (\ref{eq:Ftcoord}):
\begin{equation}
	F= \frac12 {\rm vol}_2 \ .
\end{equation}
This is similar to the ``monopole'' case in $\rr\times S^2$, which we considered in section \ref{sub:rs2mon}. Here too, this solution can be understood easily also at the spinorial level: the spin connection and gauge field combine in such a way that the CKS can be taken to be constant. 



\section{Black hole holography}
\label{sec:bh}

In this section we will consider some holographic applications of our field theory results. As already mentioned in section \ref{sub:qft} and \ref{sec: bulktoboundary}, any BPS solution in four dimensions which is AlAdS (recall that this means that the metric can be written as in (\ref{asymp})), will be related by holography to field theory on $M_3$. The geometry of $M_3$ will thus fall in our classification in sections \ref{sec:cks} and \ref{sec:coord}. Moreover, for most solutions of physical interest $M_3$ will be conformally flat, and thus will also fall in our discussion in section \ref{sec:examples}. 

In section \ref{sub:zoo} we will give some examples of such BPS solutions, both static and non-static. Many of them turn out to have naked singularities. Some, however, have a finite-area horizon: they arise in ${\cal N}=2$ gauged supergravity coupled to vector multiplets \cite{Caldarelli:1998hg,Cacciatori:2009iz,Dall'Agata:2010gj,Hristov:2010ri}. For these black hole solutions, holography can give a RG interpretation to the entire solution, as we will review in section \ref{sub:hol}. We will then sketch a conjectural discussion of this RG flow for some examples. In section \ref{subsec:twisted_sphere} we will consider black holes which have a magnetic field strength for the graviphoton, whose boundary geometry will match the one considered in section \ref{sub:rs2mon}. In section \ref{sub:torbh} we will consider black holes with a toroidal horizon, whose boundary geometry was discussed in section \ref{sub:r3}; although these solutions are physically less interesting, their field theory interpretation is perhaps clearer. Finally in section \ref{sub:hypbh} we will consider black holes with higher-genus black holes, whose boundary geometry was considered in section \ref{sub:rh2}.

\subsection{BPS solutions} 
\label{sub:zoo}

We will start by giving a quick overview of AlAdS BPS solutions in supergravity; as we remarked, their boundary geometry will then fall in our discussion in section \ref{sec:cks}. 

Let us first consider the case where the boundary geometry is $\rr\times S^2$. In the static case, as we anticipated in the introduction, there are only two classes of BPS solutions. 
\begin{enumerate}
	\item The first class consists of $1/2$ BPS solutions. They are electrically charged under the graviphoton field, which vanishes asymptotically like $1/r$. Thus the gauge field at the boundary is zero, and the CKS at the boundary is one of the CKSs discussed in section \ref{sub:rs20}. All solutions in this class known so far have naked singularities. Should any non-singular solution exist (perhaps after taking into account stringy corrections), it should be interpreted as electrically charged $1/2$ BPS states of the boundary CFT on $\rr\times S^2$.	
	
	\item The second class consists of $1/4$ BPS solutions. They are magnetically charged under the graviphoton, whose field strength goes to a constant value $F=-\frac12 {\rm vol}_2$. For this reason, their asymptotic geometry is called ``magnetic AdS'' \cite{Hristov:2011ye}. The corresponding boundary CKSs are then those discussed in section \ref{sub:rs2mon}. Holographically, these solutions correspond to deformations of the CFT on $\rr\times S^2$, as will be discussed in detail in section \ref{subsec:twisted_sphere}.	In this case, solutions with finite-area horizons have been found in \cite{Cacciatori:2009iz}, in non-minimal gauged supergravity. 
\end{enumerate}

There also exist non-static 1/4 BPS solutions in minimal gauged supergravity, given by the Kerr--Newman--AdS$_4$ black hole with some particular values of the parameters\cite{Caldarelli:1998hg}. Again, there is one solution with a vanishing magnetic field and one with a fixed non-zero magnetic charge. A suitable change of coordinates\footnote{Denoting with $\tau', \theta', \phi'$ the coordinates on the boundary that one finds in \cite{Caldarelli:1998hg}, we have \begin{equation} \tau' = \tau\ , \quad \theta' = \arctan(\sqrt{1-\alpha^2} \tan(\theta))\ , \quad \phi' = \phi + \alpha \tau\ .  \end{equation}} shows that the boundary metric of that solution is conformal to the metric (\ref{eq:rs2}) of $\rr\times S^2$. The boundary gauge field in the magnetic solution corresponds to case (3) in section \ref{sub:rs2class}, given by equation (\ref{eq:F10}). 

Finally, there are also solutions whose asymptotic geometry is $\rr\times T^2$ and $\rr\times \Sigma_g$, where $\Sigma_g$ is a Riemann surface of genus $g>1$. In the case $\rr\times T^2$, only the electrically charged solutions exist; in the case $\rr\times \Sigma_g$, only the magnetically charged solutions exist. The situation is summarized in table \ref{tab:sol}.

\begin{table}[h]\label{tab:sol}
	\begin{center}
		\setlength{\tabcolsep}{2pt}
		\begin{tabular}{ c || c | c || c | c ||} \cline{2-5}
		 & \multicolumn{2}{|c||}{AdS} &
		  \multicolumn{2}{|c||}{magnetic AdS} \vline \\ \hline
	\multicolumn{1}{|c||}{topology}	 & minimal ${\cal N}=2$ & extra matter & minimal ${\cal N}=2$ & extra matter \\ \hline
		\multicolumn{1}{|c||}{static $S^2$} & 1/2 BPS NS & 1/2 BPS NS & 1/4 BPS NS & 1/4 BPS BH \\ \hline
		\multicolumn{1}{|c||}{rotating $S^2$} & 1/4 BPS BH & ? & 1/4 BPS NS & ? \\ \hline
		\multicolumn{1}{|c||}{$T^2$} & 1/2 BPS NS & 1/4 BPS BH & $\nexists$ & $\nexists$ \\ \hline
		\multicolumn{1}{|c||}{higher genus} & $\nexists$ & $\nexists$ & 1/4 BPS BH & 1/4 BPS BH \\ \hline
		\end{tabular}
	\end{center}
	\caption{Summary of known BPS solutions. ``BH'' refers to solutions that have a finite-area horizon for at least some choice of parameters. ``NS'' refers to solutions with a naked singularity. Entries marked with a question mark have not yet been found, while $\nexists$ refers to solutions which are known not to exist.}
	\label{label}
\end{table}


\subsection{Holographic interpretation of black hole solutions} 
\label{sub:hol}

We now consider in more details  the case of static black holes with a regular horizon and we discuss their holographic interpretation. Our starting point is an asymptotically locally AdS$_4$ static BPS black hole with asymptotic behaviour as in equation (\ref{asymp}), (\ref{asympA}) and boundary metric
\begin{equation} 
  M_3=  \rr \times \Sigma_2\ , \qquad  (r \rightarrow \infty )\, ,
\end{equation}
where $\Sigma_2$  refers to $S^2$, $\mathbb{R}^2$ or a Riemann surface of genus $g>1$.
The general  BPS black holes we are interested in occur in non-minimal gauge supergravity and they have a non trivial profile for  scalar and gauge fields in vector multiplets.  According to the holographic dictionary,  the physics of these black holes is related to the physics of a three-dimensional CFT defined on $\rr \times \Sigma_2$ and explicitly deformed by the operators dual to the vector multiplet fields. These static BPS black holes, moreover, develop a near-horizon geometry AdS$_2 \times \Sigma_2$. Holographically, this should correspond to a one-dimensional theory in the IR: 
\begin{equation} 
  \rr \ , \qquad  (r \rightarrow r_{\rm hor} )\, .
\end{equation}
We expect this one-dimensional theory to be a superconformal quantum mechanics (SCQM). 
The bulk geometry therefore describes the RG flow of a three-dimensional SCFT at infinity, deformed by relevant operators and compactified on $\Sigma$, which becomes at low energies  a one-dimensional SCQM.

The theory lives on $\rr\times \Sigma_2$, but one can rewrite all its fields by their expansion on $\Sigma_2$, obtaining formally a one-dimensional theory with infinitely many fields.  The RG flow to lower energies essentially integrates out higher mass modes in the theory, and therefore it is natural to expect that the endpoint of the flow will leave only the lowest mass modes.  Given the compactness of $\Sigma_2$, this process should be equivalent to a consistent truncation down to one dimension. It would be quite interesting to perform an explicit dimensional reduction of the three-dimensional theory and study the resulting supersymmetric quantum mechanics. This reduction would be particularly interesting in examples where the three-dimensional CFT in question is known. This happens for some of the black holes in \cite{Caldarelli:1998hg,Cacciatori:2009iz,Dall'Agata:2010gj,Hristov:2010ri}: these can be uplifted to M-theory solutions\footnote{More precisely, one can uplift the four-dimensional ${\cal N}=2$ supergravity solutions with three vector multiplets (gauge group ${\rm U}(1)^4$) and a particular form of the scalar prepotential and gauging parameters \cite[Sec.~8]{Hristov:2010ri}.} which are asymptotically AdS$_4\times S^7$, and hence they correspond to deformations of the ABJM theory. The final goal of such an investigation would be the understanding and the microscopic counting of the entropy of these particular black holes.

In this paper we limit ourselves to a qualitative description of the holography for black hole solutions using supersymmetry arguments, leaving a more detailed analysis for future work. As shown in \cite[Chap.~10]{Hristov:2012bk}, each black hole is characterized by a global symmetry superalgebra, ${\cal A}_{\rm BH}$. This is the symmetry algebra of the full bulk solution, which can (and often does) get enhanced at the horizon and at infinity to  larger superalgebras ${\cal A}_{\rm hor}$ and ${\cal A}_{\infty}$:
\begin{equation}\label{constr}
	\mathcal{A}_{hor} \supseteq \mathcal{A}_{\rm BH} \subseteq \mathcal{A}_{\infty}\ .
\end{equation}
These superalgebras can be given explicitly for any particular black hole solution in supergravity/string theory.  $\mathcal{A}_{\infty}$ matches the data of the original UV CFT defined on $\rr \times \Sigma_2$   and it corresponds to the superalgebras discussed in section \ref{sub:rs2mon}, \ref{sub:r3}, and \ref{sub:rh2};  $\mathcal{A}_{\rm BH}$ captures the  symmetry of the theory which has been deformed by relevant operators;  $\mathcal{A}_{\rm hor}$ describes the low energy symmetry enhancement of the dimensionally reduced theory. In the following sections we will give the well known symmetry algebras $\mathcal{A}_{\rm hor}, \mathcal{A}_{\rm BH}, \mathcal{A}_{\infty}$ \cite[Chap.~10]{Hristov:2012bk} from the black hole side and match them by our analysis on the field theory side.


\subsection{Twisted $\rr \times S^2$: magnetic black holes} 
\label{subsec:twisted_sphere}

Our first example falls into the time-like case of section \ref{sub:rs2mon}. We consider the conformally flat spacetime $\rr \times S^2$ with a non-vanishing gauge field.

Let us first briefly describe the BPS black holes that exhibit such an asymptotic boundary. They were first found in \cite{Cacciatori:2009iz} and further discussed in \cite{Dall'Agata:2010gj,Hristov:2010ri}. Their gravity multiplet and Killing spinors precisely asymptote to the background solution in section \ref{sub:rs2mon}. The black holes develop a genuine horizon only when additional charges and running scalars are allowed: on the four-dimensional supergravity side we need at least one vector multiplet with a non-trivial scalar potential. In such a case one finds the following symmetry algebras:
\begin{equation}
	\mathcal{A}_{\rm hor}={\rm SU}(1,1|1)\times {\rm SO}(3)\ , \quad \mathcal{A}_{\rm BH} = {\rm U}(1|1)\times {\rm SO}(3)\ , \quad \mathcal{A}_{\infty}= {\rm U}(1|1)\times {\rm SO}(3)\ ,
\end{equation}
as shown in \cite{Hristov:2011ye,Hristov:2011qr,Hristov:2012nu}. It is clear that they obey \eqref{constr} and we are interested in understanding these algebras using field theory considerations.

As shown in section \ref{sub:rs2mon}, the SCFTs on the boundary have the superconformal group $\mathcal{A}_{\infty}= {\rm U}(1|1)\times {\rm SO}(3)$, as expected. From the bulk information we have, $\mathcal{A}_{\rm BH} = \mathcal{A}_{\infty}$. It follows that we need to introduce a mass deformation to bring the theory out of its UV fixed point without reducing any symmetries. This might seem unnatural, but it is exactly what happens due to the small amount of supersymmetry. Recall from (\ref{monopole-cks}) that the supersymmetry parameter for our SCFT is
\begin{equation}
\label{BH_spinor_s2}
	\epsilon= (1 + i \gamma^0) \epsilon_0\ .
\end{equation}
The bulk solution has a non trivial profile for fields belonging to vector multiplets. According to the standard AdS/CFT dictionary, we are deforming the boundary theory with  operators belonging to multiplets of currents. Since the bulk gauge fields are massless, these currents are conserved and correspond to global symmetries of our CFT. There is then a very general way of writing these deformations which is close in spirit to the way in which we defined supersymmetric field theories on curved spaces by coupling them to supergravity.  We can couple the global symmetry currents to additional gauge multiplets $\{A_{\mu}, \lambda, \lambda^{\dagger}, \sigma, D \}$ \cite{Aharony:1997bx,Fabbri:1999ay}  and treat them as background fields.  Vevs
for these background fields correspond precisely to the asymptotic values of the bulk  fields in the four-dimensional vector multiplets.  We will argue now that the supersymmetric bulk solution corresponds to the choice
\begin{equation}\label{frozen_multiplet_s2}
	F_{t \theta} = F_{t \phi} =0\ , \qquad F_{\theta \phi} = -p \sin \theta\ , \qquad \sigma=\lambda= 0\ , \qquad D = p\ ,
\end{equation}
where $p$ is here an arbitrary constant related with the value of the extra bulk magnetic charge of the black hole (which will become a dimensionful coupling constant on the field theory side). The deformation of the boundary theory is easy to describe at linear level.
There will be a  monopole background fields for  all the CFT fields charged under the global symmetries (similar to the analogous coupling for fields with a charge under the R-symmetry).  In addition there will be  mass terms that arise through the linear coupling to $D$, which schematically reads \cite{Aharony:1997bx,Fabbri:1999ay}
\begin{equation}
 	D \,  z^i T_{i\bar j} {\bar z}^{\bar j}\ ,
\end{equation}
where $z^i$ are scalar fields in the CFT and $T_{i\bar j}$ the generators of the global symmetry.   Supersymmetry is preserved since  the supersymmetry variation of the auxiliary fermions \footnote{That this is the right transformation for a theory on $\rr\times S^2$ follows by considering, for example, a four-dimensional theory on the supersymmetric background $\rr^{1,1}\times S^2$ \cite{Cassani:2012ri} and reducing the supersymmetry transformation of a vector multiplet in conformal supergravity.}
\begin{equation} \label{gaugvar}
	\delta \lambda = \left( i \gamma^{\mu} \partial_{\mu} \sigma - \frac{1}{2} F_{\mu \nu} \gamma^{\mu \nu} + i D \right) \varepsilon\ 
\end{equation}
vanishes identically if we plug in the field values \eqref{frozen_multiplet_s2} and the variation parameter \eqref{BH_spinor_s2}. 
In conclusion, the non-zero vevs of the vector mutiplet will give rise to mass terms that deform the original SCFT, thus breaking conformality without changing the supersymmetry algebra\footnote{This is only possible due to the peculiar fact that $\rr \times S^2$ with a gauge field supports ${\rm U}(1|1) \times {\rm SO}(3)$ both as a superalgebra and a superconformal algebra. This is clearly not always the case for general backgrounds.}. This concludes our proof that $\mathcal{A}_{BH} = \mathcal{A}_{\infty}$ also on the field theory side.

The last step in our prescription is to define a consistent truncation of the mass deformed three-dimensional theory on $S^2$ to one dimension. We leave the explicit details of this for a future investigation, noting that this is not an usual sphere compactification due to the fact that the fermionic fields of our theory are twisted, and become scalars under rotation in this theory; we saw this happen for the supersymmetry parameter $\epsilon$ (see our discussion around \eqref{monopole-cks}). This matches precisely the bulk expectations \cite{Hristov:2011ye}. The resulting quantum mechanics in general would have a ${\rm U}(1|1) \times {\rm SO}(3)$ symmetry. The theory will depend on the parameter $p$ which appeared in (\ref{frozen_multiplet_s2}); for particular values of $p$ it should become exactly conformal, and the time translation become part of the larger group ${\rm SO}(2,1) \cong {\rm SU}(1,1)$. This should enhance the symmetry group to ${\rm U}(1|1) \rightarrow {\rm SU}(1,1|1)$ (i.e.~replacing ${\rm U}(1) \rightarrow {\rm SU}(1,1)$), thus recovering the expected horizon symmetry algebra, $\mathcal{A}_{hor}$.


\subsection{$\rr \times T^2$: toroidal black holes} 
\label{sub:torbh}

Our second example is in a way easier to understand, as it is based on the very well known maximally conformally symmetric solution on $\rr^{1,2}$. Recall from section \ref{sub:r3} that a CKS in Minkowski$_3$ space-time with a vanishing gauge field reads
\begin{equation}
	\varepsilon = \varepsilon'_0 - (t \gamma_0 + x \gamma_1 + y \gamma_2) \varepsilon''_0\ , 
\end{equation}
for arbitrary constant spinors $\varepsilon'_0, \varepsilon''_0$. Since we are interested in compact horizons, we would like to periodically identify the Cartesian coordinates $x \sim x+1, y \sim y+1$ to become circular directions $T^2$. This imposes $\varepsilon''_0=0$, and we are left with four conserved supercharges from $\varepsilon'_0$. We can also take the original ten CKVs on Minkowski and check which ones are periodic in $x, y$. This leaves us only with the translations $\partial_t, \partial_x, \partial_y$ and thus breaks ${\rm SO}(3,2)$ to $\rr \times {\rm U}(1)^2$.\footnote{For some particular values of $\tau$, there are also finite isometries, but we will assume in what follows that $\tau$ is generic.} Together with the ${\rm U}(1)_{\rm R}$-symmetry, these CKSs and CKVs generate the superconformal algebra on $\rr \times T^2$, which is non-semisimple and was already given in \cite{Hristov:2012bd} as the superalgebra of toroidal black holes\footnote{At this stage the bulk and boundary calculations of $\mathcal{A}_{\infty}$ coincide due to the fact that both follow from ${\rm OSp}(2|4)$. This is however still an independent field theory check due to the different starting points in three and four dimensions.}.

Going briefly to the bulk picture, there exist 1/4 BPS black holes with regular horizons only when extra vector multiplets with magnetic charges and running scalars are added \cite{Cacciatori:2009iz,Hristov:2012bd}, in analogy to the spherical case of the previous subsection. The relevant symmetry algebras are\footnote{We already matched the non-semisimple $\mathcal{A}_{\infty}$ and thus leave it out of further discussions.}:
\begin{equation}
	  \mathcal{A}_{BH} = {\rm U}(1|1)\times {\rm U}(1)^2\ , \qquad \mathcal{A}_{hor}={\rm SU}(1,1|1)\times {\rm U}(1)^2\ .
\end{equation}
	  Unlike the case of section \ref{subsec:twisted_sphere}, here we see from $\mathcal{A}_{BH}$ that the mass deformed theory is expected to break the original four fermionic symmetries to only two. Even though the way to introduce the mass deformation here is similar to (\ref{frozen_multiplet_s2}), namely:
\begin{equation}
		F_{t x} = F_{t y} =0\ , \qquad F_{xy} = -p \ , \qquad \sigma=\lambda= 0\ , \qquad D = p\ 
\end{equation}	
  (due to the same type of bulk solutions for the vector multiplets of spherical and toroidal black holes \cite{Cacciatori:2009iz}), the gaugino variation (\ref{gaugvar}) is not  automatically satisfied for the constant CKS $\varepsilon'_0$. We need the additional projection
\begin{equation}
	\varepsilon' = (1 + i \gamma^0) \varepsilon'_0 
\end{equation}	  
	  to make sure that $\delta \lambda = 0$. This leaves the mass deformed theory with the expected symmetry group ${\rm U}(1|1)\times {\rm U}(1)^2$ 
\cite{Hristov:2012bd}.
	  
	  The last step of consistent truncation on $T^2$ is more standard than for the case of a sphere, but we still do not perform it explicitly here. Whether the final supersymmetric quantum mechanics is conformal will again depend crucially on the value of the mass deformation parameter $p$ (the charge in the bulk picture). Just as before, using the argument that one-dimensional conformal invariance enhances ${\rm U}(1|1)$ to ${\rm SU}(1,1|1)$ we recover the correct near-horizon symmetry group $\mathcal{A}_{\rm hor}$.


\subsection{$\rr \times H_2 / S_n$: higher genus black holes} 
\label{sub:hypbh}

Let us now consider the spacetime $\rr \times H_2 / S_n$, where we take the quotient of the hyperbolic space by a suitable discrete subgroup of its isometry group ${\rm SO}(2,1)$ (see \cite{Caldarelli:1998hg} for the same construction on the bulk side). This describes all higher genus $g > 1$ Riemann surfaces. Therefore this spacetime is related to the possibility of different horizon topologies in asymptotically AdS$_4$ black holes \cite{Caldarelli:1998hg,Cacciatori:2009iz}. The isometry group ${\rm SO}(2,1)$ will be broken by the quotient procedure to a finite subgroup, and generically to the identity. We will assume here we are in the generic case.   

As a starting point, we consider the CKS solution on $\rr \times H_2$ coming from the time translation vector discussed below \eqref{notimportant} in section \ref{sub:rh2}. We are again in a situation with a constant magnetic field on the Riemann surface and a CKS solution
\begin{equation}\label{hyp_twisted}
	\varepsilon = (1 + i \gamma^0) \varepsilon_0 \ .
\end{equation}
This spinor does not depend on the spatial coordinates and therefore survives all quotients of $H_2$. Note that this is only valid for the particular choice of solution on $H_2$ with a magnetic field stregth, while the other allowed CKSs from section \ref{sec:examples} would not respect the identifications and thus not lead to a SCFT on the compact surface. This is of course not a coincidence, since the same is true on the bulk side. We know that there exist BPS black holes with higher genus surface horizons only for this choice of asymptotic boundary \cite{Caldarelli:1998hg,Cacciatori:2009iz,Hristov:2012bk}. From this point on, the discussion becomes analogous to the one in section \ref{subsec:twisted_sphere}. The relevant symmetry algebras on the black hole side are
\begin{equation}
	\begin{split}
	&\mathcal{A}_{hor}={\rm SU}(1,1|1)\ , \\  
	&\mathcal{A}_{BH} = {\rm U}(1|1)\ ,\\ 
	&\mathcal{A}_{\infty}= {\rm U}(1|1)\ .	
	\end{split}
\end{equation}
This should match with the field theory analysis following the same steps as in section \ref{subsec:twisted_sphere}: the mass deformation does not break any additional symmetries, while the reduction potentially enhances ${\rm U}(1|1)$ to ${\rm SU}(1,1|1)$.

\section*{Acknowledgments}
We would like to thank J.~Figueroa O'Farrill, N.~Halmagyi, S.~Katmadas, C.~Klare and D.~Klemm for interesting discussions. 
The authors are supported in part by INFN, by the MIUR-FIRB grant RBFR10QS5J ``String Theory and Fundamental Interactions'', and by the MIUR-PRIN contract 2009-KHZKRX. The research of A.T.~is also supported by the ERC Starting Grant 307286 (XD-STRING).

\appendix

\section{Conformal structure} 
\label{sec:conf}

A general field configuration $\{ \phi_i \}$ is said to have a conformal symmetry along a vector $k$ if every field $\phi_i$ satisfies 
\begin{equation}\label{CKV_lie_der}
	{\cal L}_k \phi_i = \lambda_i \phi_i\ ,
\end{equation} 
where ${\cal L}_k$ is the Lie derivative, and $\lambda_i$ are some scalar functions. Of particular importance for us is to consider a spacetime with a conformal isometry. A conformal Killing vector (CKV) for the metric $g_{\mu \nu}$ satisfies (in local coordinates)
\begin{equation}\label{CKV}
	({\cal L}_k g)_{\mu\nu}=\nabla_{\mu} k_{\nu} + \nabla_{\nu} k_{\mu} = \lambda g_{\mu \nu}\ ,
\end{equation} 
where $\nabla$ is the spacetime covariant derivative. Every isometry of the metric is also a conformal isometry, or in other words all Killing vectors are also conformal Killing vectors, but not vice versa. Under a Weyl rescaling 
\begin{equation}\label{eq:weyl}
	g_{\mu\nu} \to e^f g_{\mu\nu}
\end{equation}
a CKV remains a CKV, without any rescaling:
\begin{equation}\label{eq:zweyl}
	z \to z \ .
\end{equation}
Therefore a spacetime can be characterized in terms of its conformal isometries only up to conformal rescalings.

The fermionic equivalent of a CKV is the conformal Killing spinor (CKS) $\epsilon$, which in $d$ dimensions is a solution of the equation
\begin{equation}\label{intro:CKS}
	\left(\nabla_\mu - \frac{1}{d} \gamma_\mu (\gamma^{\nu} \nabla_{\nu}) \right) \epsilon = 0\ ;
\end{equation} 
as we have seen in the main text, $\nabla$ is often replaced by a gauge-covariant $\nabla^A=\nabla-i A$. Under Weyl rescaling (\ref{eq:weyl}), from a CKS $\epsilon$ one can obtain a CKS of the rescaled metric by 
\begin{equation}\label{eq:epsweyl}
	\epsilon \to e^{f/4} \epsilon \ .
\end{equation}

The number of independent solutions to equations \eqref{CKV} (or more generally \eqref{CKV_lie_der}) and \eqref{intro:CKS} tells us the number of bosonic and fermionic conformal isometries, respectively. (In section \ref{sec:cks} we will see that these two equations are intimately related.) One can further construct the full superconformal symmetry algebra by the commutations relations between CKVs and CKSs. Note however that both equations are conformally covariant: under conformal rescaling of the metric, the conformal isometries are invariant. Thus the superconformal symmetries remain the same for two different metrics related by a conformal transformation. 

We will now consider some examples. We focus on four dimensions, which is the case most readers will be more familiar with. Similar considerations apply to the three-dimensional case.

The most common example is when we consider four-dimensional Minkowski and $\mathbb{R} \times S^3$ spacetimes. They have different isometries (Poincar\'{e} vs. $\mathbb{R}\times {\rm SO}(3)$) and one is maximally supersymmetric in ungauged supergravities, while the other does not admit any Killing spinors. However, due to their conformal equivalence, they share the exact same conformal (${\rm SO}(4,2)$) and superconformal symmetries ${\rm SU}(2,2|{\cal N})$. One can also think of both these spacetimes as being possible boundary foliations of AdS$_5$, which explains why we find the same symmetries for the resulting SCFTs. There are in fact infinitely many spacetimes that are conformally equivalent to both Minkowski and $\mathbb{R} \times S^3$ in four dimensions, since one can always consider an arbitrary conformal rescaling. 

One should however be slightly more careful with arbitrary examples --- note that \eqref{CKV} and \eqref{intro:CKS} are local equations, i.e.~they do not carry information about the global spacetime structure. This can in principle give rise to subtleties. For example, after the conformal rescaling the spacetime might be no longer geodesically complete, and it is a priori not obvious that one can continue the CKV to the analytically continued spacetime. (We have seen this for Minkowski$_3$ and $\rr\times S^2$ in section \ref{sub:r3}.) Another possible subtlety arises from global identifications. For example, in section \ref{sec:bh} we encountered a superconformal field theory on $\mathbb{R} \times T^2$ and saw that it has fewer globally defined symmetries than flat space, even though their local structure is the same. We would therefore like to stress that a given result from the CKS and CKV equations immediately holds for the full class of conformally equivalent metrics, but always up to global issues that need to be considered case by case.

\subsection{Example: SCFTs on de Sitter} 
\label{sub:dS}

Let us treat in more detail a particularly interesting example, namely de Sitter spacetime.
We will see that this space admits the same type of superconformal field theories as Minkowski space. 

Once again, although in the main text we consider three-dimensional theories, let us consider here four-dimensional de Sitter spacetime, dS$_4$, which might be more familiar to the reader. It is well-known that its symmetry group is ${\rm SO}(4,1)$ and that it does not admit any Killing spinors, i.e.~it cannot have fermionic symmetries. This was conclusively shown by \cite{Pilch:1984aw} since the bosonic algebra ${\rm SO}(4,1)$ does not have a unitary fermionic completion. Therefore one can certainly expect the supersymmetric field theories on dS$_4$ to be non-unitary and thus unphysical. This is in accordance with the fact that usual field theories on de Sitter acquire a natural (non-zero) temperature, related to the Hubble radius \cite{Gibbons:1977mu}. Such thermal theories were more recently considered from the point of view of the AdS/CFT correspondence in \cite{Marolf:2010tg}.  

If one is considering a conformal theory, however, the discussion changes. After all, de Sitter represents an expanding universe, but a conformal theory should be insensitive to overall rescalings of distances. dS$_4$ is conformally flat, and the earlier discussion in this section then shows that it has conformal group ${\rm SO}(4,2)$, and conformal supergroup ${\rm SU}(2,2|{\cal N})$. Explicitly, one can use the fact that global dS$_4$ (with metric $ds^2= - d\tilde t^2 + \cosh^2(\tilde t) ds^2_{S^3}$) is conformal to the region $-\pi/2 < \tau < \pi/2$ of $\rr \times S^3$ via the change of coordinates $\cosh(\tilde t)= \frac 1{\cos(\tau)}$, and then use the transformation laws (\ref{eq:zweyl}), (\ref{eq:epsweyl}) to transport the CKVs and CKSs of $\rr\times S^3$ \footnote{The CKSs of $\rr\times S^3$ can be written similarly to the CKSs of $\rr\times S^2$ in (\ref{eq:CKSrs2}).} to dS$_4$. Another possibility is to work in the planar patch, where the metric reads $\frac1{\eta^2}(-d \eta^2 + d \vec x^2)$, and is hence conformal to one-half of Minkowski space.

It is then clear that supersymmetry on de Sitter is intimately related with conformal symmetry. One can formulate a SCFT field theory on dS$_4$ by coupling the theory to conformal supergravity (as discussed in \cite{Klare:2012gn} in four dimensions; we saw  in section \ref{sub:qft} the analogue of that discussion for three dimensions), but not a more general supersymmetric theory. Conversely, a breaking of the conformal isometries immediately leads to a supersymmetry breaking. 

As a last remark, note that the same discussion holds equally well for more general types of cosmological FLRW metrics (as well as for negative curvature spacetimes like AdS \cite{Aharony:2010ay}), up to possible global issues as explained above.





\providecommand{\href}[2]{#2}
\end{document}